\documentclass[11pt,a4paper]{article}

\usepackage{lmodern} 
\usepackage[T1]{fontenc}
\usepackage[utf8]{inputenc}
\usepackage{amsmath,amsthm,amssymb}
\usepackage{natbib}          
\usepackage{graphicx}        
\usepackage{tikz}
\usepackage{url}
\usepackage{float}           
\usepackage[capitalise]{cleveref}
\usepackage{microtype}       
\usepackage[a4paper,top=2.5cm,bottom=2.5cm,left=3cm,right=3cm]{geometry}  
\newcommand{\claims}{\mathbf{c}}
\newcommand{\cost}{\mathrm{Cost}}
\DeclareMathOperator{\dep}{dep}
\newcommand{\lo}{\underline{g}}
\newcommand{\hi}{\overline{g}}

\newtheorem{theorem}{Theorem}[section]
\newtheorem{proposition}[theorem]{Proposition}
\newtheorem{lemma}[theorem]{Lemma}
\newtheorem{corollary}[theorem]{Corollary}
\theoremstyle{definition}
\newtheorem{definition}[theorem]{Definition}

\usepackage{xcolor}

\begin{document}

\title{Bankruptcy Rules for Strategic Agents}
\author{Yosef Zumer, Erel Segal-Halevi, David Sarne}
\date{\today}
\maketitle
\begin{abstract}
Bankruptcy rules are rules for dividing an insufficient estate among people with different claims. 
Existing rules assume that the claims are publicly known. But when this is not the case, the claimants have a strong incentive to claim more than they are due, as all bankruptcy rules give larger allocations to claimants with larger claims.
A remedy often used by legal systems is to make over-reporting costly by requiring evidence and imposing penalties, but this remedy also increases the legal expenses of proving true claims, which might result in under-reporting. Encouraging true reporting requires fine-tuning of the legal costs.

This paper is the first to incorporate costs of supporting both true and false claims into the bankruptcy model.
Claimants incur costs for the true and false parts of their claims, with false claims costing at least as much per unit as true claims.
The goal is to set cost rates that induce claimants to report their true entitlements, neither more nor less.
Whether this is possible at a positive true-claim cost rate depends on the allocation rule.
The proportional and constrained equal awards rules cannot guarantee truthfulness at any positive true-claim cost rate;
the same holds for constrained equal losses with at least three claimants.

For $N\ge2$ claimants, the rule TAL--$1/N$ from the parameterized Talmud family induces truthfulness exactly when
the true-claim cost rate is at most $1/N$ and the false-claim cost rate is at least $1-1/N$.
This rule is optimal in the sense that its truthful cost region contains that of every piecewise smooth allocation rule: it gives the legal system the maximum possible flexibility in setting the legal costs.
For two claimants, TAL--$1/2$ (also called the Contested Garment rule) uniquely attains this optimal region.

Because honest claimants bear the cost of proving their claims, truthfulness also consumes part of the estate; we show that the rules attaining the largest truthful region are also those that maximize social welfare subject to truthfulness.
\end{abstract}

\section{Introduction}\label{sec:intro}
The \emph{bankruptcy problem} is one of the oldest fair allocation problem, tracing back to early appearances in the
Talmud and in the writings of Maimonides.
It involves several \emph{claimants}, each of whom has a monetary \emph{claim}, and a fixed \emph{estate} that is insufficient to satisfy all claims fully.
The challenge is to design a rule for allocating the estate among the claimants.

The most widely used rule in practice is the \emph{proportional rule}, by which each claimant receives an amount proportional to its claim. Other common rules are  \emph{constrained equal awards (CEA)} (each claimant receives the minimum between its claim and a fixed amount) and
\emph{constrained equal losses (CEL)} (each claimant loses the minimum between its claim and a fixed amount). The Talmud rule, mentioned in the Babylonian Talmud, applies a particular combination of CEA and CEL.

In an attempt to decide which rule is the fairest of them all, many researchers have characterized these and other rules  axiomatically (as the unique rule that satisfies a set of natural fairness and consistency axioms) and strategically (as the equilibrium of some natural game based on the claims); 
See \citet{thomson2019how} for a textbook survey of the topic.

An important assumption made by almost all works on the bankruptcy problem is that all claims are given \emph{exogenously}.
In practice, however, claimants have an obvious incentive to claim more than they are due. This is because all allocation rules  satisfy a natural fairness requirement called \emph{claim monotonicity} --- increasing a claimant's claim never decreases his award, and may increase it. 
There is an inherent tension between fairness and incentives.

Legal systems try to avoid these tension by making false claims \emph{costly}. This can be done by requiring evidence to substantiate the claims, investigating the evidence in court, and imposing high penalties once false claims are caught.
A false-claim cost higher than its marginal gain would prevent rational claimants from inflating their claims.
However, these measures make truthful claims costly too, as they require truthful claimants to collect evidence, withstand investigation, and pay lawyers.
Although the false-claim cost is usually higher (e.g. due to the possible penalty), the true-claim cost is still non-negligible.
A true-claim cost higher than its marginal gain would prevent rational claimants from claiming their true entitlement, which is as undesirable as a false claim.
This raises the following primary question:
\begin{quote}
(Q1) \emph{Which bankruptcy rules, and which false and true claim costs, incentivize all claimants to truthfully report their claims?}
\end{quote}

Moreover, as the costs cannot always be set with full precision, one would like the allocation rule to be as flexible as possible with respect to the allowed costs. This raises the following secondary question:
\begin{quote}
(Q2) \emph{Which bankruptcy rules incentivize all claimants to truthfully report their claims, under the widest range of false and true claim costs?}
\end{quote}

\paragraph{Contribution.}
We formally extend the classic bankruptcy model to handle incentives under litigation costs.
In our model, claim\-ants may report any amount up to the estate.
Much like all work on the bankruptcy problem, our focus is on the \emph{conflict domain}, where the sum of claims is strictly larger than the estate (in the non-conflict domain all claimants receive exactly their claims, so all bankruptcy rules are identical and the above two questions are not meaningful).

Each report splits into a \emph{true claim}, up to the claimant's entitlement,
and a \emph{false claim} beyond it.

Each claimant incurs costs that are linear in the claimed amounts, reflecting that lawyers’ fees are commonly stated as a percentage of the disputed amount and that fraud penalties typically rise with the amount involved.


Both cost rates are nonnegative. The false-claim cost rate is at least as high as the true-claim cost rate.
The relation between the rates may be any nondecreasing function:
deterring false claims may also make honest proof more expensive.

\textbf{(Q1): A criterion for truthfulness.}
A rule is truthful exactly when
the true-claim cost rate is at most its smallest marginal gain and the false-claim cost rate is at least its largest marginal gain (\cref{prop:criterion}).
The first bound makes honest proof worthwhile; the second deters over-reporting.
Based on this principle, we prove that the proportional rule, CEA and CEL (for three or more claimants), are truthful if and only if the true-claim cost rate is $0$ and the false-claim cost rate is at least $1$ (\cref{thm:prop,thm:cea,thm:cel}).

\textbf{(Q2): Optimal allocation rules.}
For two claimants, the Contested Garment (CG) rule \citep{aumann1985game} is truthful whenever the true cost rate is at most $1/2$ and the false cost rate is at least $1/2$.
This is provably the widest possible truthful cost range: for any piecewise-smooth allocation rule, the truthful cost range is contained in the range of CG (\Cref{cor:cg-optimal}).
This makes CG the optimal rule with respect to incentives.

For any $N\geq 2$ claimants, an allocation rule known as \emph{TAL--$1/N$} \citep{moreno2006tal} supports truthful reporting whenever the true cost rate is at most $1/N$ and the false cost rate is at least $(N-1)/N$; this is provably the widest range of claim costs among all piecewise smooth allocation rules (\cref{thm:taln}).
Note that for two claimants, TAL--$1/2$ coincides with CG.

\textbf{Weak and strict truthfulness.}
The standard definition of truthfulness requires that the truthful report is a best response.
In this paper we call it \emph{weak truthfulness}.
\emph{Strict truthfulness} makes the truthful report the \emph{unique} best response.
We extend some of our results to strict truthfulness in \Cref{sec:strict}.

The proportional rule is strictly truthful exactly when the true-cost rate is $0$ and the false-cost rate is at least $1$, whereas CEA and CEL (for three or more claimants) are never strictly truthful.
TAL--$1/N$ is strictly truthful exactly when the true-claim cost rate is less than $1/N$ and the false-claim cost rate is more than $1-1/N$; these bounds also describe CG when $N=2$ (\cref{prop:strict-taln,prop:strict-cg-cel}).

Finally, we ask what is the cost of truthfulness. Honest claimants still pay to prove their claims, so litigation consumes part of the estate even when nobody lies. \Cref{sec:welfare} measures this loss and shows what governs it: the court must set the false-claim rate high enough to deter the most profitable lie the rule permits anywhere, and since the two rates are linked, honest claimants pay for that deterrence in every case. 
The rules with the largest truthful region turn out to be among those that minimize this loss, so TAL–$1/N$ maximizes social welfare subject to truthfulness (\Cref{cor:tal-welfare}). Incentives and efficiency do not conflict here: the same rule is best on both counts.
  
Two messages follow. First, the division rule and the court’s evidentiary regime cannot be designed in isolation from one another. A court that fixes how much proof to demand without regard to the rule that will divide the estate may destroy truthful reporting altogether, and the rule most widely used in practice, the proportional rule, is precisely the one that fails for any positive cost of honest proof. Second, the rule that does survive is not a new construction. TAL-$1/N$ reduces for two claimants to the Contested Garment rule recorded in the Talmud. A division principle formulated two millennia ago turns out to be optimal once the cost of honesty is taken into account, uniquely so for two claimants.


\newpage
\section{Related Work}\label{sec:related}\par

The textbook by \citet{thomson2019how} provides a comprehensive survey of the research related to the bankruptcy problem. Below we present the works more closely related to ours.

\subsection{Classical and Talmudic Allocation Rules}

\citet{oneill1982problem} introduced the formal model, together with the properties of the
\emph{proportional rule} that avoid manipulation.
This rule is the one most used in practice.
Two other classical rules are widely studied: the
constrained equal awards rule (CEA), rooted in the writings of Maimonides \citep{aumann1985game},
protects small claimants by equalizing the amounts received, while the constrained equal losses
rule (CEL) equalizes the losses, the amounts by which the awards fall short of the claims, and so
favors large claimants. 

An important milestone was the publication of the paper by \citet{aumann1985game},
who were motivated by \citet{oneill1982problem},
which analyzed a famous Talmudic case concerning an insufficient estate left by a man who married
three women. Aumann and Maschler demonstrated that the specific
allocation prescribed by the Talmud, known as the \emph{TAL rule}, which had puzzled scholars
for nearly two thousand years, is consistent with another division rule from the Talmud, the
\emph{Contested Garment rule (CG)}: every two claimants divide their combined award exactly
as CG prescribes (\Cref{sec:rules}).
The TAL rule moves between the CEA and CEL principles as the estate grows.

\citet{moreno2006tal} then generalized
it into the parameterized TAL family: a continuous family of rules whose endpoints are exactly
CEA and CEL, and whose middle member is the TAL rule itself. \citet{moreno2004securement}
characterized the TAL rule by bilateral consistency, self-duality, and \emph{securement}, which
gives every claimant at least the smaller of the claim and the estate, divided by the number of
claimants.

All these characterizations take claims as exogenously given.

\subsection{Strategic Manipulation and Non-Manipulable Rules}
Truthfulness is an important concern in many mechanism design and social choice settings.
However, in the context of the bankruptcy problem it has not received much attention, as truthfulness seems to contradict the most basic property of \emph{claim monotonicity} (higher claims win higher shares).

One kind of manipulation that has been studied is the splitting of a claim into two separate claims, or its opposite --- merging two claims into one. 
\citet{oneill1982problem} introduced the requirement, now known as \emph{no advantageous merging or splitting}, that pooling several claims or dividing one among new claimants (possibly fictitious) never changes the total award, and proved that together with symmetry, continuity, efficiency, and independence of dummy claimants it characterizes the \emph{proportional rule}.
\citet{defrutos1999coalitional} separated the two directions of the requirement and studied
non-manipulability by merging, by splitting, and by both.
\citet[Theorem~2]{ju2003manipulation} characterized non-manipulability by pairwise splitting
and by pairwise merging under equal treatment of equals, consistency, and continuity.
\citet{ju2007nonmanipulable} generalized the model, letting a claim be
multi-dimensional, and characterized the rules under which no group of claimants can raise the
total amount it receives by transferring claims within the group. \citet{moreno2007manipulation}
extended the question to the parameterized TAL family.
A member of the TAL family is non-manipulable by merging
when the estate's share of total claims is below its parameter,
and by splitting when it is above. At equality both properties hold.
CEA deters merging and CEL deters splitting on the whole domain.
The paper also ranks these rules by their relative non-manipulability.

\subsection{Strategic Choice of Division Rules}
Another kind of strategic consideration in bankruptcy is the choice of the rule that should divide the estate.
\citet{chun1989noncooperative} studied a related surplus-sharing problem in which agents
choose allocation rules satisfying efficiency, fair ranking, and progressivity.
Theorem~1 of that paper states that every pure-strategy Nash equilibrium of its procedure
gives equal shares of the surplus.
For bankruptcy problems, \citet{herrero2003equal} used dual procedures for resolving
disagreements to give noncooperative support to CEA and CEL.

\citet{bouwhuis2026strategic} restricted the available rules to the parameterized TAL family.
Each claimant chooses a parameter between zero and one,
and the rule at the mean of the chosen parameters divides the estate.
The game has a Nash equilibrium for every bankruptcy problem.
For two claimants, all equilibria yield the TAL allocation, whose parameter is one half.

\subsection{Claim Games and Strategic Justifications}

Axiomatic characterizations select rules by their properties; strategic justifications derive
their allocations as equilibrium outcomes.
In the claim game of \citet[Section~5]{oneill1982problem}, each claimant places a fixed total claim
on chosen parts of the estate, and each part is divided equally among those who claim it.
Equilibrium allocations need not be unique.
\citet{peters2019claim} generalized the integer claim game and showed
that when no claim exceeds the estate and the estate is at least half of the total of the claims,
every way of dividing the contested parts in full that treats equal claims equally
yields the same set of equilibrium payoffs.
The TAL rule selects a payoff in this set on this domain.

In a bargaining approach, \citet{sertel1992nash} studied two-person problems with a zero
disagreement point and a feasible set that is a right triangle with positive ideal payoffs.
The optimal pre-donation by the player with the larger ideal payoff makes the Nash bargaining
solution on the resulting feasible set yield the Contested Garment allocation, when ideal payoffs
are interpreted as claims and the estate equals the larger claim.
\citet{serrano1995strategic} uniquely implemented the TAL allocation,
the nucleolus of the associated cooperative game \citep{aumann1985game}, in subgame-perfect
equilibrium through bilateral negotiations.
\citet{dagan1997noncooperative} gave a sequential procedure whose unique Nash equilibrium
outcome is the allocation prescribed by any consistent, estate-monotone, and supermodular rule.
Later bargaining procedures justified the proportional, CEA, CEL, and TAL rules within one
framework \citep{tsay2019relations}, and then the whole parameterized TAL family
\citep{moreno2022strategic}.

In the multilateral bargaining game of \citet[Theorem~1]{hagiwara2021strategic},
the unique pure-strategy subgame-perfect equilibrium outcome is the CEA allocation in the first period.
Three divide-and-choose games yield CEA, CEL, and TAL, respectively, as unique subgame-perfect
equilibrium outcomes \citep{li2023divide}.
These results justify allocations through strategic interaction with given claims.

\citet{dagan1999feasible} implement every taxation method in Nash, strong, and coalition-proof
equilibrium through reports of all agents' incomes, with partial verification ruling out overstatements.
Agents know all incomes; the planner does not.

\citet{ashlagi2012estate} let the claimants state the claims
themselves and applied the rule to what they stated: when the claimants share one upper bound on
what they may claim, and the rule is efficient, treats equal claimants equally, and preserves the
order of the awards, every pure-strategy equilibrium gives each claimant the smaller of that bound
and an equal share of the estate. Free claiming thus erases the differences between the rules in
equilibrium, and that paper closes by leaving open the case in which the claimants spend resources
to support their claims.

\subsection{Costly Claims and Verification}
Verification of claims has been studied in domains other than bankruptcy.

\emph{Costly verification models} study how a designer can induce truthful reporting when checking
private information is expensive.
\citet{townsend1979costly} studied lending contracts with costly verification of a borrower's return;
\citet{border1987samurai} jointly chose payments and audit probabilities to deter wealth under-reporting.
For allocation without monetary transfers, \citet{benporath2014optimal} derived an optimal
mechanism that induces truthful reporting
when the principal pays for verification and chooses allocation and verification probabilities.
In strategic classification, \citet{estornell2021incentivizing} studied
the choice of an audit policy under a limited audit budget, with fines for agents caught misreporting their features to obtain resources.
These models make verification part of the mechanism design problem.

Misreporting can also be costly without an audit.
\citet{kartik2009strategic} studied communication with a direct cost of misrepresenting private
information, showing how lying costs affect information revelation.
In his quadratic lying-cost example, the cost depends on the deviation from the truth and vanishes at truthful communication.

These studies examine how costs affect incentives for truthful reporting.
To the best of our knowledge, truthful reporting that involves costs of supporting a claim in the bankruptcy problem has not been studied.
In this paper, we study bankruptcy allocation rules when supporting claims is costly
and characterize the conditions under which they induce truthful reporting.

\newpage

\section{Model} \label{sec:model}

This section recalls the standard bankruptcy problem, the properties an allocation rule may satisfy,
and the classical rules on which the later analysis rests. It then extends that model to strategic
reporting and to litigation costs.

\subsection{The Standard Bankruptcy Problem}

A bankruptcy problem \citep{oneill1982problem} is a tuple $(E, N, \claims)$
consisting of an \emph{estate} $E>0$, a number of \emph{claimants} $N\ge2$ labeled by $[N] := \{1,
\ldots, N\}$, and a vector of \emph{claims} $\claims = (c_1, \ldots, c_N) \in \mathbb{R}_{+}^{N}$.

Write $C = \sum_{i \in [N]} c_i$ for the sum of the claims, $c_{-i}$ for the claims of all
claimants other than $i$, and $C_{-i} := \sum_{j\ne i} c_j = C - c_i$ for their sum; as $N$ is
implicit in the dimension of $\claims$, the problem is described as $(E, \claims)$.

\subsection{Allocation Rules and Basic Properties}

An allocation rule is a mapping
\[
\phi : \mathbb{R}_{+} \times \mathbb{R}_{+}^{N} \to \mathbb{R}_{+}^{N}
\]
assigning to each problem $(E, \claims)$ an allocation $\phi(E, \claims) = (\phi_i(E,
\claims))_{i\in [N]}$. Feasibility requires that no claimant receive more than his claim,
\[
0 \le \phi_i(E, \claims) \le c_i \quad \text{for every } i \in [N],
\]
and that the estate not be over-allocated,
\[
\sum_{i \in [N]} \phi_i(E, \claims) \le E.
\]

The analysis relies on the following standard properties. A rule $\phi$ is:
\begin{itemize}
 \item \emph{efficient} if it allocates the entire estate whenever the claims exceed it,
and pays all claims otherwise:
 \[
\sum_{i \in [N]} \phi_i(E, \claims) = \min\{E,C\}.
\]
In particular, on the conflict domain $C>E$,
 \[
\sum_{i \in [N]} \phi_i(E, \claims) = E.
\]

 \item \emph{monotone} if raising a claimant's own claim never lowers his allocation; that is, for
$c_i \le c'_i$ with $c_{-i}$ fixed,
 \[
 \phi_i\bigl(E,(c_i,c_{-i})\bigr) \le \phi_i\bigl(E,(c'_i,c_{-i})\bigr);
 \]
  \item \emph{continuous} if for every $i$ and every $(E,c_{-i})$ the award $\phi_i(E,\claims)$ is
continuous in claimant $i$'s own claim $c_i\in[0,E]$.

 \item \emph{piecewise smooth} if the rule is continuous and for every $i$ and every $(E,c_{-i})$
the award $\phi_i(E,\claims)$ is continuously differentiable except in finitely many points, and
the derivative has finite one-sided limits everywhere, so a one-sided slope exists at every claim.
\end{itemize}
Every rule of \cref{sec:rules} qualifies. CEA, CEL, CG, and the
TAL--$\theta$ family are piecewise \emph{linear}, and the proportional rule is smooth. Every rule is
also efficient and monotone throughout, monotonicity making all marginal gains
non-negative. 

On the non-conflict domain $C\leq E$, we have $\phi(E,\claims)=\claims$ for any rule $\phi$.
At $C=E$ efficiency and feasibility force $\phi_j=c_j$ for every $j$, so each award
agrees with the full payment made outside the conflict domain.

\subsection{Common Rules}\label{sec:rules}

The analysis covers the following rules.
\begin{itemize}
\item \emph{Proportional rule}:
\[
\phi^{\text{PROP}}_i(E, \claims) = \frac{c_i}{C}\, E.
\]
\item \emph{Constrained Equal Awards} (CEA):
\[\phi^{\text{CEA}}_i(E, \claims) = \min(c_i,\, \alpha),\quad \alpha \ge 0 \text{ solving }
\textstyle\sum_{j \in [N]} \min(c_j, \alpha) = E.\]
\item \emph{Constrained Equal Losses} (CEL):
\[\phi^{\text{CEL}}_i(E, \claims) = \max(0, c_i - \beta),\quad \beta \ge 0 \text{ solving }
\textstyle\sum_{j \in [N]} \max(0, c_j - \beta) = E.\]
\item \emph{Contested Garment} (CG) rule (two claimants):
\[\phi^{\text{CG}}_i\!\bigl(E,\,(c_i, c_j)\bigr) = \frac{E + \min(c_i, E) - \min(c_j, E)}{2}. \]
\item \emph{TAL rule}:
\[
\phi^{\text{TAL}}_i(E, \claims) =
\begin{cases}
\phi^{\text{CEA}}_i\!\left(E,\;\tfrac{1}{2}\claims\right) & \text{if } E \leq \tfrac{1}{2}C, \\[2mm]
\tfrac{1}{2}c_i + \phi^{\text{CEL}}_i\!\left(E - \tfrac{1}{2}C,\;\tfrac{1}{2}\claims\right) &
\text{if } E > \tfrac{1}{2}C.
\end{cases}
\]
Equivalently, TAL is the unique rule consistent with CG \citep{aumann1985game}: for $x =
\phi^{\text{TAL}}(E, \claims)$ and any pair $i,j$, $x_i = \phi_i^{\text{CG}}(x_i + x_j,\, (c_i,
c_j))$.
\item \emph{TAL--$\theta$ rule} (for $\theta \in [0,1]$), the family of \citet{moreno2006tal}:
\[
\phi^{\text{TAL-}\theta}_i(E, \claims) =
\begin{cases}
\phi^{\text{CEA}}_i(E,\, \theta\claims) & \text{if } E \leq \theta C, \\[2mm]
\theta c_i + \phi^{\text{CEL}}_i\!\left(E - \theta C,\, (1-\theta)\claims\right) & \text{if } E >
\theta C,
\end{cases}
\]
or equivalently, in closed form,
\[
\phi^{\text{TAL-}\theta}_i(E, \claims) =
\begin{cases}
\min(\theta c_i,\, \alpha) & \text{if } E \leq \theta C, \\[2mm]
\max(\theta c_i,\, c_i - \beta) & \text{if } E > \theta C,
\end{cases}
\]
where $\alpha$ solves $\sum_{j}\min(\theta c_j,\alpha)=E$ and $\beta$ solves $\sum_{j}\max(\theta
c_j,\,c_j-\beta)=E$.\footnote{In each defining equation the left-hand side is continuous and
monotone in the parameter, non-decreasing in $\alpha$ and non-increasing in $\beta$, and takes a
value $\le E$ at one end of the parameter range and a value $\ge E$ at the other, so a solution
exists. The induced allocation is unique even when the defining equation determines the parameter
only up to an interval of solutions. The same remark covers $\alpha$ in CEA and $\beta$ in CEL.} The
TAL rule is the special case $\theta = \tfrac{1}{2}$; the endpoints $\theta=1$ and $\theta=0$
reproduce CEA and CEL respectively.
\end{itemize}

\subsection{Marginal Gains and Kinks}

The central object of the analysis is claimant $i$'s \emph{marginal gain}: the rate at which his
allocation responds to his own claim. The direction of a misreport matters, so the two
\emph{one-sided derivatives} are kept apart:
\begin{align*}
\begin{aligned}
g_i^{+}(\phi, E,\claims) &:= \lim_{\varepsilon \to 0^{+}}
  \frac{\phi_i(E,\, c_i + \varepsilon,\, c_{-i}) - \phi_i(E,\claims)}{\varepsilon},\\
g_i^{-}(\phi, E,\claims) &:= \lim_{\varepsilon \to 0^{+}}
  \frac{\phi_i(E,\claims) - \phi_i(E,\, c_i - \varepsilon,\, c_{-i})}{\varepsilon}.
\end{aligned}
\end{align*}
That is, $g_i^{+}$ is the award won per unit of over-reporting and $g_i^{-}$ the award lost per unit
of under-reporting. By piecewise smoothness both exist and are finite at every interior point (at
the endpoints only the inward derivative is defined: $g_i^{+}$ at $c_i=0$, $g_i^{-}$ at $c_i=E$).

A \emph{kink} is a claim at which $g_i^+$ and $g_i^-$ disagree.

For CEA, the gain falls from $1$ to $0$ at $c_i=\alpha$; for CEL, it rises from $0$
at $c_i=\beta$ (\Cref{fig:kinks}).

\subsection{Strategic Reporting}\label{sec:reporting}

Each claimant $i \in [N]$ may report a claim $c_i$ differing from his entitlement $c_i^*$. By
assumption, reports (and entitlements) are restricted to $[0,E]$: the court does not allow
claims exceeding the estate.

Write $\claims^*=(c_1^*,\ldots,c_N^*)$ for the true entitlement profile and $C^*:=\sum_{i\in[N]}c_i^*$ for its total.

Any report decomposes into a \emph{true claim} and a \emph{false
claim},
\[
c_i^T := \min(c_i, c_i^*), \qquad c_i^F := c_i - c_i^T = \max(0,\, c_i - c_i^*),
\]
so that $c_i = c_i^T + c_i^F$: under-reporting ($c_i \le c_i^*$) gives $c_i^T = c_i,\ c_i^F = 0$,
and over-reporting ($c_i > c_i^*$) gives $c_i^T = c_i^*,\ c_i^F = c_i - c_i^*$.

\subsection{Strategic Litigation Costs and the Cost Dependence}\label{sec:dependence}

A false claim costs at least as much per unit as a true one, and cost scales with the claim's size.
Taking both rates to be constant, claimant $i$'s litigation cost is
\[
\cost_i(c_i;\,c_i^*) = p^T \cdot c^T_i + p^F \cdot c^F_i,
\qquad p^F \ge p^T \ge 0,
\]
and his net utility from the profile of reports is
\begin{equation}\label{eq:utility}
U_i(E, \claims) = \phi_i(E, \claims) - \cost_i(c_i;\,c_i^*).
\end{equation}

\paragraph{Dependence between $p^T$ and $p^F$.}
The two rates are dependent because the true and false parts of a report
are examined in the same legal process.
Within legal and institutional constraints, the court sets $p^F$ through
evidence requirements, investigations, and fines for false reporting.
Making false claims more costly can therefore also increase the cost rate $p^T$
of establishing true claims.
A higher $p^T$ harms honest claimants and the social welfare as we will see in \cref{sec:welfare}.

This dependence is represented by a given nondecreasing function
$\dep:\mathbb{R}_{+}\to\mathbb{R}_{+}$:
\[
p^T=\dep(p^F).
\]

\subsection{The Conflict Domain}\label{sec:conflict}
Classical bankruptcy research takes claims as given and assumes that they are in the \emph{conflict domain},
where the estate cannot meet all claims \citep{oneill1982problem,aumann1985game,moreno2006tal}.
The strategic model studied here distinguishes reported claims from true entitlements (\cref{sec:reporting}),
so the analysis assumes that true entitlements are in the conflict domain, that is, $C^*>E$.

For fixed $E$ and $c_{-i}$, define claimant $i$'s \emph{conflict domain} by
\[
\mathcal D_i(E,c_{-i}):=\{x\in[0,E]:x+C_{-i}>E\}.
\]
Here $C^*_{-i}:=\sum_{j\ne i}c_j^*$.
Thus, $c^*_i\in\mathcal D_i(E,c^*_{-i})$ exactly when $c^*_i\in[0,E]$ and the profile $(c^*_i,c^*_{-i})$ is in conflict.


The assumption $C^*>E$ is equivalent to $c_i^*\in\mathcal D_i(E,c^*_{-i})$ for every $i\in[N]$.

The restriction to the conflict domain is motivated by the fact that if $c_i^*\not\in\mathcal D_i(E,c^*_{-i})$,
then claimant $i$ receives the entire claim $c_i$ for \emph{any} $c_i\not\in\mathcal D_i(E,c^*_{-i})$.
Hence, claimant $i$ has a strong incentive to report the highest (or supremum) value that is not in $\mathcal D_i(E,c^*_{-i})$.
This incentive is common to all allocation rules and cannot be avoided.
Hence, the analysis is restricted to the conflict domain, where rules differ in their handling of these incentives.

In the following analysis of a claimant's incentives, we assume w.l.o.g. that the other claimants report their true entitlements, as the current claimant's incentives are not affected by whether or not the others are truthful.
For fixed $E$, $i$, and $c^*_{-i}$, varying $c_i$ gives $\claims=(c_i,c^*_{-i})$ and $C=c_i+C^*_{-i}$.

\newpage
\section{Incentives for Truthful Reporting}\label{sec:incentives}

This section studies when reporting the true claim is a best response for every claimant.
\Cref{sec:criterion} defines this notion of truthfulness and characterizes exactly which cost rates
guarantee it: the answer depends on the rule through two values only, its smallest and its largest
marginal gain. \Cref{sec:robust} then asks which dependence between the two rates allows truthful
reporting, and \cref{sec:master} illustrates the answer with a linear dependence.

Throughout, a \emph{conflict profile} denotes a pair $(E,\claims)$ in the conflict domain $C>E$ with
every claim in $[0,E]$; both conditions are implicit.

\subsection{Truthfulness}\label{sec:criterion}

With $E$, $c_{-i}$, and $c_i^*$ fixed, the net utility \eqref{eq:utility} becomes a function of
claimant $i$'s own claim alone, $U_i(c_i):=U_i\bigl(E,(c_i,c_{-i})\bigr)$.

\begin{definition}[Truthfulness]\label{def:truthful}
Fix cost rates $(p^T,p^F)$. A rule $\phi$ is \emph{$(p^T,p^F)$-truthful} if,
for every $E>0$, $i\in[N]$, $c_{-i}\in[0,E]^{N-1}$ and
$c_i^*\in\mathcal D_i(E,c_{-i})$,
truthful reporting is a best response: $U_i(c_i^*)\ge U_i(c_i)$
for every report $c_i$.
\end{definition}

This is the weak form: other reports may tie with the truthful one. Throughout
\cref{sec:incentives,sec:classical}, truthful stands for this weak notion. The strict form, under
which truthful reporting is the unique best response, is treated separately in \cref{sec:strict}.

Starting from the true claim, a marginal reduction saves cost at rate $p^T$
and reduces the award at rate $g_i^-$.
If $p^T>g_i^-$, the claimant saves more than he loses, so under-reporting is profitable.
Preventing this at every conflict profile requires $p^T\le\inf g_i^-$.

A marginal increase costs $p^F$ per unit and increases the award at rate $g_i^+$.
If $g_i^+>p^F$, the claimant gains more than he pays, so over-reporting is profitable.
Preventing this at every conflict profile requires $p^F\ge\sup g_i^+$.
For each claimant $i$, define
\[
\lo_i(\phi) := \inf g_i^-(\phi, E, \claims),\qquad \hi_i(\phi) := \sup g_i^+(\phi, E, \claims),
\]
where the infimum and supremum range over all conflict profiles $(E,\claims)$
at which the respective one-sided derivative is defined.

The extrema over the claimants define the \emph{extremal marginal gains}
\[
\lo(\phi):=\min_i \lo_i(\phi),
\qquad
\hi(\phi):=\max_i \hi_i(\phi).
\]
The argument $(\phi)$ is omitted when the rule is understood.
Throughout, $\hi<\infty$ is assumed; every rule of \cref{sec:rules} satisfies $\hi\le1$.

\begin{samepage}
\begin{lemma}[Bounds on the marginal gains]\label{lem:dispersion}
Let $\phi$ be a piecewise smooth allocation rule.
Then
\[
\sum_{i\in [N]} \lo_i(\phi) \le 1,
\qquad
\sum_{i\in [N]} \hi_i(\phi) \ge N-1,
\]
so $\lo\le\tfrac1N$ and $\hi\ge\tfrac{N-1}{N}$.
In particular, if a rule has constant marginal gain $\mu = \lo = \hi$, then $\mu=\tfrac1N$.
\end{lemma}
\end{samepage}
\begin{proof}
Fix $x\in[E/(N-1),E]$ and write $\mathbf{x}=(x,\dots,x)$.
With $c_{-i}=\mathbf{x}_{-i}$, every own claim $t\in(0,x]$ gives
$C=t+(N-1)x>E$.
Feasibility gives $\phi_i(E,(0,\mathbf{x}_{-i}))=0$.

By piecewise smoothness, there are finitely many kinks $k_1<\cdots<k_m$ in $(0,x)$.
Set $k_0=0$ and $k_{m+1}=x$.
For every $j=0,\ldots,m$ and $t\in(k_j,k_{j+1})$, the function $\phi_i(E,(t,\mathbf{x}_{-i}))$ is continuously
differentiable and $t\in\mathcal D_i(E,\mathbf{x}_{-i})$, so
\[
\lo_i\le\frac{d}{dt}\phi_i(E,(t,\mathbf{x}_{-i}))
=g_i^-(\phi,E,(t,\mathbf{x}_{-i}))
=g_i^+(\phi,E,(t,\mathbf{x}_{-i}))\le\hi_i.
\]
Continuity allows the integral to be split at these kinks, giving
\[
\phi_i(E,\mathbf{x})
=\sum_{j=0}^{m}\int_{k_j}^{k_{j+1}}\frac{d}{dt}\phi_i(E,(t,\mathbf{x}_{-i}))\,dt
\in[\lo_i x,\hi_i x].
\]
Summing and using efficiency gives
\[
x\sum_i\lo_i\le E\le x\sum_i\hi_i.
\]
Taking $x=E$ in the left inequality and $x=E/(N-1)$ in the right proves the two sum bounds.
Taking the minimum and maximum over claimants gives
$\lo\le\tfrac1N\sum_i\lo_i\le\tfrac1N$ and
$\hi\ge\tfrac1N\sum_i\hi_i\ge\tfrac{N-1}{N}$.
If $\lo=\hi=\mu$, these bounds imply
$\mu\le\tfrac1N\le\tfrac{N-1}{N}\le\mu$, so $\mu=\tfrac1N$.
\end{proof}

\begin{proposition}[Truthful region]\label{prop:criterion}
Let $\phi$ be a piecewise-smooth allocation rule. Then $\phi$
is $(p^T,p^F)$-truthful if and only if $p^T\le\lo(\phi)$ and $p^F\ge\hi(\phi)$.
\end{proposition}
\begin{proof}

\emph{Necessity.} Suppose $p^T>\lo(\phi)$. Then some claimant $i$ has $\lo_i(\phi) < p^T$.
Then some conflict profile $E,\claims$ has
$g_i^-(\phi,E,\claims)<p^T$.

By piecewise smoothness, there exists
$0<\delta<\min\{c_i,C-E\}$ so that $(c_i-\delta,c_i)$ contains no kinks.
As $t$ increases from $c_i-\delta$ to $c_i$, the derivative $\frac{d}{dt}\phi_i(E,(t,c_{-i}))$ tends to $g_i^-(\phi,E,\claims)<p^T$, hence
some $\eta>0$ satisfies
$\frac{d}{dt}\phi_i(E,(t,c_{-i}))\le p^T-\eta$ for every $t\in(c_i-\delta,c_i)$.

Set $\claims^*=\claims$, so $C^*=C>E$ and $c_i^*\in\mathcal D_i(E,c^*_{-i})$.
For $0<\varepsilon<\delta$, absolute continuity gives
$\phi_i(E,(c_i^*,c_{-i}))-\phi_i(E,(c_i^*-\varepsilon,c_{-i}))$\allowbreak${}=\int_{c_i^*-\varepsilon}^{c_i^*}\frac{d}{dt}\phi_i(E,(t,c_{-i}))\,dt$\allowbreak${}\le(p^T-\eta)\varepsilon$, so
\[
\begin{aligned}
U_i(c_i^*-\varepsilon)-U_i(c_i^*)
&=p^T\varepsilon-\bigl[\phi_i(E,(c_i^*,c_{-i}))-\phi_i(E,(c_i^*-\varepsilon,c_{-i}))\bigr]\\
&\ge p^T\varepsilon-(p^T-\eta)\varepsilon=\eta\varepsilon>0,
\end{aligned}
\]
which incentivizes under-reporting; hence $p^T\le\lo(\phi)$. 

Symmetrically, if $p^F<\hi(\phi)$ there are a conflict
profile and a claimant with $g_i^+(\phi)>p^F$ and $c_i<E$;
setting $\claims^*=\claims$, a sufficiently small
over-report is strictly profitable; hence $p^F\ge\hi(\phi)$.

\emph{Sufficiency.} $p^T\le\lo(\phi)$ and $p^F\ge\hi(\phi)$: then at every conflict profile
$g_i^-(\phi,E,\claims)\ge\lo(\phi)\ge p^T$ and $g_i^+(\phi,E,\claims)\le\hi(\phi)\le p^F$.
Fix $E$, $c_{-i}$, and $c_i^*\in\mathcal D_i(E,c_{-i})$.

\emph{Over-reporting ($c_i>c_i^*$).}
Since $c_i^*\in\mathcal D_i(E,c_{-i})$, every $t\in(c_i^*,c_i)$ belongs to
$\mathcal D_i(E,c_{-i})$ too,
where $\frac{d}{dt}\phi_i(E,(t,c_{-i}))-p^F$\allowbreak$=g_i^+(\phi,E,(t,c_{-i}))-p^F\le0$ at every differentiable claim.

Let $k_1<\cdots<k_m$ be the kinks in $(c_i^*,c_i)$,
and set $k_0=c_i^*$ and $k_{m+1}=c_i$.
For every $j=0,\ldots,m$, $(k_j,k_{j+1})\subseteq\mathcal D_i(E,c_{-i})$.
At every $t\in(k_j,k_{j+1})$,
\[
\frac{d}{dt}\phi_i(E,(t,c_{-i}))=g_i^+(\phi,E,(t,c_{-i}))\le\hi\le p^F.
\]
Splitting the integral at these kinks gives
\[
\begin{aligned}
U_i(c_i)-U_i(c_i^*)
&=\int_{c_i^*}^{c_i}\bigl(\frac{d}{dt}\phi_i(E,(t,c_{-i}))-p^F\bigr)\,dt
\\
&
=\sum_{j=0}^{m}\int_{k_j}^{k_{j+1}}\bigl(\frac{d}{dt}\phi_i(E,(t,c_{-i}))-p^F\bigr)\,dt\le0.
\end{aligned}
\]
hence $U_i(c_i)\le U_i(c_i^*)$.

\emph{Under-reporting ($c_i<c_i^*$).} 

If $c_i<E-C_{-i}$, the interval from $c_i$ to $c_i^*$ includes
a full-payment part, $[c_i,E-C_{-i}]$, followed by a conflict part, $(E-C_{-i},c_i^*]$.
If $c_i\ge E-C_{-i}$,
then
the first interval is empty and
$(c_i,c_i^*]\subseteq\mathcal D_i(E,c_{-i})$.

To cover both cases, set $a:=\max\{c_i,E-C_{-i}\}$.
Thus $c_i\le a<c_i^*$, since $c_i<c_i^*$ and $E-C_{-i}<c_i^*$.
In both cases, $\phi_i(E,(a,c_{-i}))-\phi_i(E,(c_i,c_{-i}))=a-c_i$, with both sides zero when $a=c_i$.

Let $k_1<\cdots<k_m$ be the kinks in $(a,c_i^*)$,
and set $k_0=a$ and $k_{m+1}=c_i^*$.
For $j=0,\ldots,m$, $\phi_i(E,(t,c_{-i}))$ is continuously differentiable as a function of $t$ on $(k_j,k_{j+1})$.
Every $t\in(k_j,k_{j+1})$ satisfies $t>a\ge E-C_{-i}$,
so $t\in\mathcal D_i(E,c_{-i})$ and
\[
\frac{d}{dt}\phi_i(E,(t,c_{-i}))=g_i^-(\phi,E,(t,c_{-i}))\ge\lo\ge p^T.
\]
Integrating gives
\[
\begin{aligned}
U_i(c_i^*)-U_i(c_i)
&
=\int_{c_i}^{c_i^*}\bigl(\frac{d}{dt}\phi_i(E,(t,c_{-i}))-p^T\bigr)\,dt
\\
&
=(1-p^T)(a-c_i)
+\sum_{j=0}^{m}\int_{k_j}^{k_{j+1}}\bigl(\frac{d}{dt}\phi_i(E,(t,c_{-i}))-p^T\bigr)\,dt\ge0.
\end{aligned}
\]
Here $1-p^T>0$ because $p^T\le\lo\le1/N<1$ by \cref{lem:dispersion}.
\end{proof}

\begin{corollary}[Bounds on truthful cost rates]\label{cor:pt-bound}
Let $\phi$ be piecewise smooth. If $\phi$ is $(p^T,p^F)$-truthful, then
\[
p^T\le\tfrac1N,
\qquad
p^F\ge\tfrac{N-1}{N}.
\]
\end{corollary}

\begin{proof}
By \cref{prop:criterion}, truthfulness requires $p^T\le\lo$ and $p^F\ge\hi$.
The two bounds follow from \cref{lem:dispersion}.
\end{proof}

\begin{definition}[Optimal rules]\label{def:optimal}
Fix $N\geq 2$. A piecewise smooth allocation rule is \emph{optimal} if it is
$(\tfrac1N,1-\tfrac1N)$-truthful.
The class of such rules is denoted by $\mathrm{OPT}_N$.
\end{definition}

By \cref{prop:criterion,lem:dispersion}, $\phi\in\mathrm{OPT}_N$ if and only if
\begin{equation}\label{eq:optimal-gains}
\lo=\tfrac1N,\qquad \hi=1-\tfrac1N.
\end{equation}
Its truthful region is therefore exactly
\begin{equation}\label{eq:optimal-region}
0\le p^T\le\tfrac1N,\qquad p^F\ge1-\tfrac1N.
\end{equation}
By \cref{cor:pt-bound}, this region contains the truthful region of every piecewise smooth rule
for the same number of claimants.
Thus optimality means having the largest truthful cost region.
\Cref{thm:taln} below shows that $\mathrm{OPT}_N$ is nonempty for any $N\geq 2$.

\paragraph{Equal cost rates.}
For $N\ge3$, \cref{cor:pt-bound} rules out truthfulness at equal cost rates,
since $\tfrac{N-1}{N}>\tfrac1N$.
For $N=2$, \cref{prop:criterion,lem:dispersion} give truthfulness at equal cost rates if and only if
\[
p^T=p^F=\lo=\hi=\tfrac12.
\]

\subsection{Truthfulness under Cost Dependence}\label{sec:robust}

When $p^T$ and $p^F$ are set independently, a truthful pair always exists: by \cref{prop:criterion}
the corner $(p^T,p^F)=(0,\hi)$ is truthful for every piecewise smooth rule. The question is whether
truthfulness relies on honest proof being free.

For a nondecreasing dependence $p^T=\dep(p^F)$, the
rule $\phi$ is \emph{truthful under $\dep$} if it is
$(\dep(p^F),p^F)$-truthful for some $p^F\ge\dep(p^F)$.

\begin{proposition}[Truthfulness under cost dependence]\label{prop:robust}
A piecewise smooth rule is truthful under $\dep$ if and only if
\[
\dep(\hi)\le\lo.
\]
\end{proposition}

\begin{proof}
\emph{Necessity.} If a truthful pair exists, \cref{prop:criterion} gives
$\dep(p^F)\le\lo$ and $p^F\ge\hi$.
Since $\dep$ is nondecreasing, $\dep(\hi)\le\dep(p^F)\le\lo$.

\emph{Sufficiency.} If $\dep(\hi)\le\lo$, choose $p^F=\hi$.
Then $p^T=\dep(\hi)\le\lo\le\hi=p^F$ by \cref{lem:dispersion},
so the cost rates satisfy the model restriction and induce truthfulness by \cref{prop:criterion}.
\end{proof}

When $\lo=0$, truthfulness requires honest proof to be free, $p^T=0$
(\Cref{prop:criterion}).
Under a nondecreasing dependence, a truthful pair then exists if and only if $\dep(\hi)=0$.
Hence $\dep(\hi)>0$ rules out truthfulness.
When $\lo>0$, truthfulness can hold even with $p^T>0$.

\newpage
\subsection{The Linear Cost Dependence}\label{sec:master}
As a simple illustration of the general criterion in \cref{prop:robust}, 
we consider the case that the true cost rate is a linear function of false cost rate, $p^T = \dep(p^F)=\lambda p^F$, with fixed $\lambda\in[0,1]$,
Here $\lambda$ is the ratio of the two cost rates when $p^F>0$. $\lambda=0$ means that honest proof is free; $\lambda=1$ means that the cost rates are equal.

\begin{corollary}[Linear cost dependence]\label{cor:master}
For any allocation rule $\phi$, 
let $\lambda^{*}(\phi):={\lo(\phi)}/{\hi(\phi)}$.
In the illustrative linear-dependence model with ratio $\lambda$, a piecewise smooth rule $\phi$ is truthful if and only if
\begin{equation}\label{eq:lambda-star}
\lambda^{*}(\phi)
\ge
\lambda.
\end{equation}
The bound $\lambda^{*}(\phi)$ is well defined since $\hi\ge\tfrac{N-1}{N}>0$
(\Cref{lem:dispersion}).
Thus the truthful range of $\lambda$ is $[0,\lambda^{*}(\phi)]$.
For these values of $\lambda$, the truthful values of $p^F$ are $[\hi,\infty)$ when
$\lambda=0$ and $\bigl[\hi,\lo/\lambda\bigr]$ when $\lambda>0$.
\end{corollary}

\begin{proof}
Since $\lambda\in[0,1]$ and $\hi>0$, the pair $(\lambda\hi,\hi)$ satisfies $p^T\le p^F$.
By \cref{prop:robust} and monotonicity, truthfulness holds if and only if
$\lambda\hi\le\lo$; dividing by $\hi>0$ gives \eqref{eq:lambda-star}.
By \cref{prop:criterion}, the remaining condition is $p^F\ge\hi$ when $\lambda=0$;
for $\lambda>0$, the condition $\lambda p^F\le\lo$ also gives $p^F\le\lo/\lambda$.
\end{proof}

If $\lo=0$, then $\lambda^{*}=0$.
Thus truthfulness under linear dependence is possible only at $\lambda=0$, with $p^F\ge\hi$.

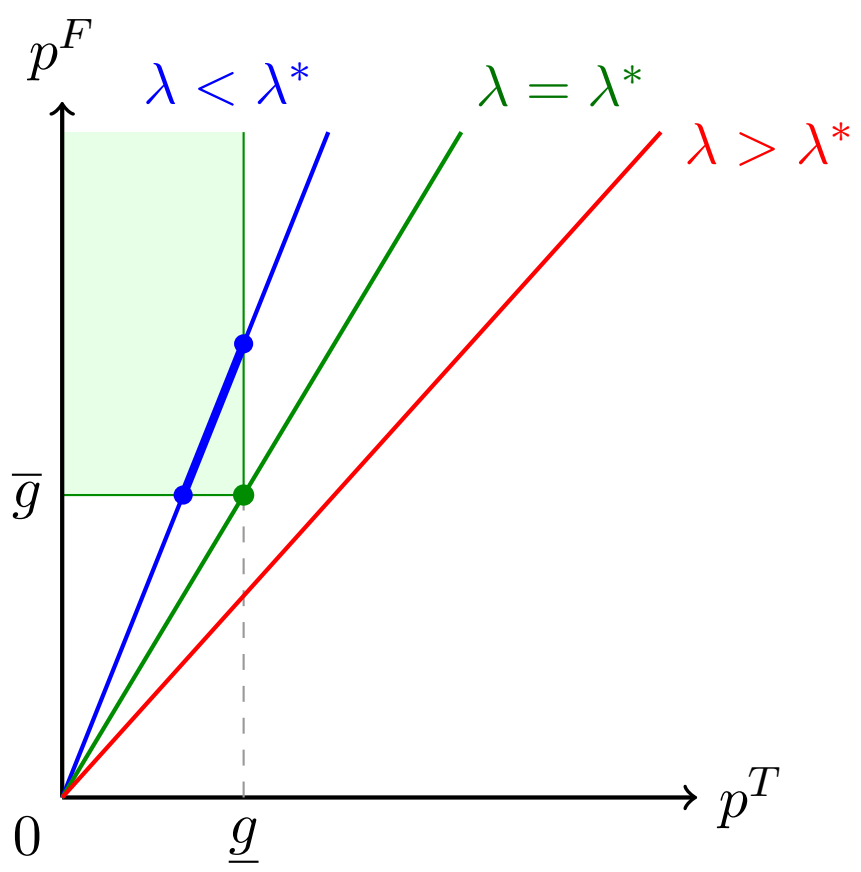
\begin{figure}[H]
\centering
\begin{tikzpicture}
\fill[green!10] (0,2) rectangle (1.2,4.4);
\draw[green!55!black] (1.2,2) -- (1.2,4.4);
\draw[green!55!black] (0,2) -- (1.2,2);
\draw[->, thick] (0,0) -- (4.2,0) node[right] {$p^T$};
\draw[->, thick] (0,0) -- (0,4.6) node[above] {$p^F$};
\node[below left] at (0,0) {$0$};
\draw[black!40, dashed] (1.2,0) -- (1.2,2);
\node[below] at (1.2,0) {$\lo$};
\node[left] at (0,2) {$\hi$};
\draw[thick, blue] (0,0) -- (1.76,4.4);
\draw[ultra thick, blue] (0.8,2) -- (1.2,3);
\filldraw[blue] (0.8,2) circle (1.6pt);
\filldraw[blue] (1.2,3) circle (1.6pt);
\node[blue, above left] at (1.8,4.44) {$\lambda<\lambda^{*}$};
\draw[thick, green!55!black] (0,0) -- (2.64,4.4);
\filldraw[green!55!black] (1.2,2) circle (1.8pt);
\node[green!45!black, above right] at (2.6,4.44) {$\lambda=\lambda^{*}$};
\draw[thick, red] (0,0) -- (3.96,4.4);
\node[red, right] at (3.98,4.32) {$\lambda>\lambda^{*}$};
\end{tikzpicture}
\caption{The truthful region $p^T\le\lo$, $p^F\ge\hi$ (green) and the lines $p^T=\lambda p^F$ below,
at, and above the bound $\lambda^{*}=\lo/\hi$.
At $\lambda=0$, the line is the axis $p^T=0$.}
\label{fig:master}
\end{figure}

\Cref{fig:master} illustrates \cref{cor:master}.

\newpage
\section{The Classical Rules}\label{sec:classical}
The preceding section reduced the question of truthfulness to two values:
the smallest and largest marginal gains, $\lo$ and $\hi$.
These determine which cost pairs make truthful reporting a best response
and whether a given cost dependence permits such a pair (\cref{prop:criterion,prop:robust}).
The next step is therefore to compute these gains for common allocation rules.

The calculations first show that the proportional and CEA rules, and CEL for $N\ge3$, have $\lo=0$.
Thus, these rules can be truthful only when reporting true claims is costless.
This motivates the search for a rule that allows truthfulness when true reporting is costly.
For two claimants, the Contested Garment rule does so: its marginal gain is always $1/2$ for $c_i\in\mathcal D_i(E,c_{-i})$.
The analysis then shows that it is the unique piecewise smooth rule with the largest truthful cost region.

For three or more claimants, a common constant marginal gain is impossible.
The final part therefore turns to the TAL family, seeking a rule whose smallest gain is as large as possible
and whose largest gain is as small as possible.
TAL--$1/N$ attains both universal bounds and hence has a truthful cost region containing
that of every piecewise smooth rule for the same $N$.

\subsection{The Proportional Rule}\label{sec:prop}

\begin{theorem}
\label{thm:prop}
The proportional rule has $\lo=0$ and $\hi=1$
Consequently it is $(p^T,p^F)$-truthful if and only if $p^T=0$ and $p^F\ge1$.
The rule is untruthful under every nondecreasing dependence with $\dep(1)>0$.
For illustration, under linear dependence it is truthful only at $\lambda=0$.
\end{theorem}

\begin{proof}
The allocation in the conflict domain ($C>E$) is $\phi_i = \tfrac{c_i}{c_i + C_{-i}} E$;
it is smooth in $c_i$, with
\[
g_i = \frac{\partial \phi_i}{\partial c_i} = \frac{E \cdot C_{-i}}{(c_i +
C_{-i})^2}=\frac{E}{C}\cdot\frac{C_{-i}}{C}.
\]

Since $C_{-i}=C-c_i>E-c_i\ge0$, the gain satisfies $0<g_i\le E/C<1$.

For $0<\varepsilon<E/2$, the profiles $c_i=E$, $C_{-i}=\varepsilon$ and
$c_i=2\varepsilon$, $C_{-i}=E-\varepsilon$ both have $C=E+\varepsilon$.
Their gains tend to $0$ and $1$, respectively, as $\varepsilon\to0^+$.
Thus $\lo=0$ and $\hi=1$, and neither value occurs as a marginal gain.

\emph{Dependence.} The truthfulness claims follow from \cref{prop:criterion,prop:robust},
and \cref{cor:master} gives $\lambda^{*}=0$.
\end{proof}

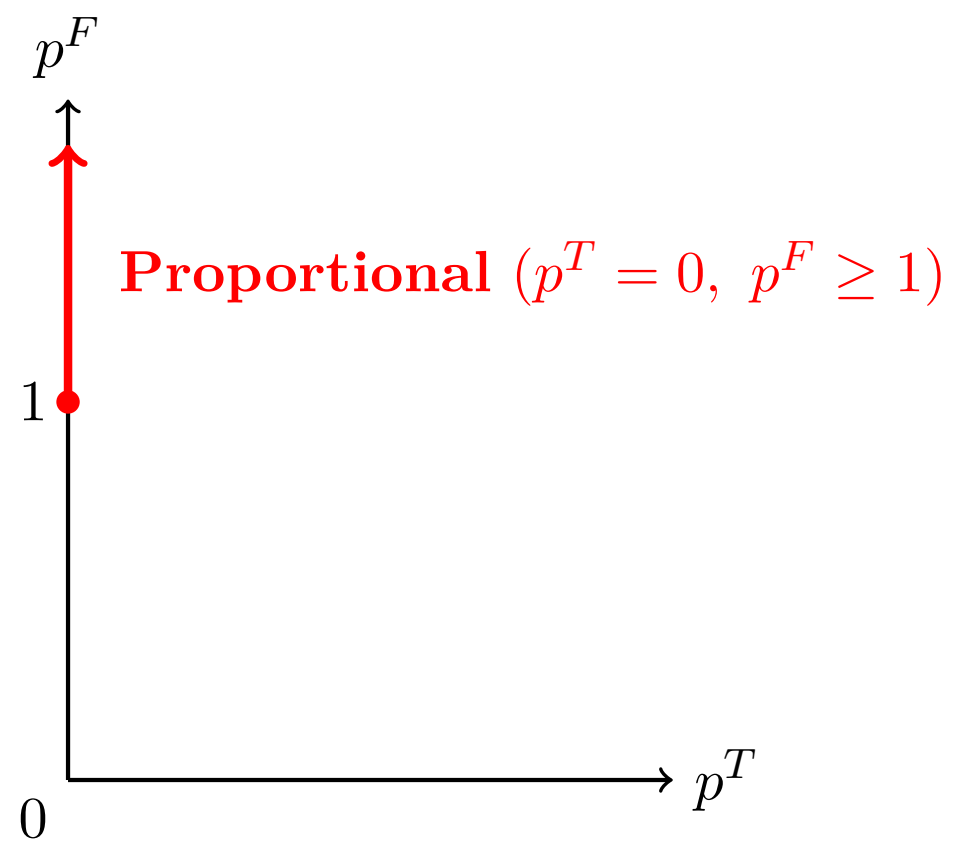
\begin{figure}[H]
\centering
\begin{tikzpicture}
 \draw[->, thick] (0,0) -- (4,0) node[right] {$p^T$};
 \draw[->, thick] (0,0) -- (0,4.5) node[above] {$p^F$};
 \node[below left] at (0,0) {$0$};
 \draw[ultra thick, red, ->] (0,2.5) -- (0,4.2);
 \filldraw[red] (0,2.5) circle (2pt) node[left, text=black] {$1$};
 \node[red, right] at (0.2, 3.35) {\textbf{Proportional} ($p^T=0,\ p^F \ge 1$)};
\end{tikzpicture}
\caption{Truthful region for the proportional rule: $p^T=0,\ p^F\ge1$ (\Cref{thm:prop}).}
\label{fig:prop}
\end{figure}

\Cref{fig:prop} shows the truthful region.

\begin{samepage}
\subsection{The Constrained Equal Awards Rule}\label{sec:cealcel}
Denote by $S_{\mathrm{CEA}}$ the claimants receiving $c_j$ under CEA, and by $S_{\mathrm{CEA}}^c$ its complement:
\begin{align*}
    S_{\mathrm{CEA}} &= \{j: \phi^{CEA}_j = c_j\}
    \\
    S_{\mathrm{CEA}}^c &= \{j: \phi^{CEA}_j < c_j\}.
\end{align*}

\begin{lemma}[Finitely many changes of side under CEA]\label{lem:cea-sides}
Fix $E$ and $c_{-i}$.
As $c_i$ varies,
the sets $S_{\mathrm{CEA}}$ and $S_{\mathrm{CEA}}^c$ change at finitely many points.
\end{lemma}
\end{samepage}
\begin{proof}
For $C\le E$, the sets are constant.
For $C>E$, the set $S_{\mathrm{CEA}}^c$ is nonempty.
If $\alpha$ increased as $c_i$ increased, no award $\min(c_j,\alpha)$ would decrease,
and every claimant initially in $S_{\mathrm{CEA}}^c$ would receive more.
This would contradict efficiency, so $\alpha$ is nonincreasing in $c_i$.

For $c_i'>c_i$, let $\alpha'$ denote the corresponding parameter, so $\alpha'\le\alpha$.
Since each rival's claim $c_j$ is fixed,
\[
\begin{aligned}
(c_i'-\alpha')-(c_i-\alpha)&=(c_i'-c_i)+(\alpha-\alpha')>0,\\
(c_j-\alpha')-(c_j-\alpha)&=\alpha-\alpha'\ge0 \qquad (j\ne i).
\end{aligned}
\]
Hence $c_i-\alpha$ is strictly increasing, and each rival's $c_j-\alpha$ is nondecreasing.
A claimant $j$ belongs to $S_{\mathrm{CEA}}^c$ exactly when $c_j-\alpha>0$.
Once positive, this difference stays positive as $c_i$ increases.
Each claimant therefore changes sets at most once, proving finiteness.
\end{proof}

\begin{lemma}[CEA is piecewise linear and continuous]\label{lem:cea-pl}
Fix $E$ and $c_{-i}$.
The award $\phi^{\text{CEA}}_i$ is continuous and piecewise linear in $c_i$.
\end{lemma}
\begin{proof}
By \cref{lem:cea-sides}, the sets are fixed between finitely many claims.
On each such interval with $C>E$, $S_{\mathrm{CEA}}^c$ is nonempty, and efficiency gives
\[
\alpha=\frac{E-\sum_{j\in S_{\mathrm{CEA}}}c_j}{|S_{\mathrm{CEA}}^c|}.
\]
The awards $c_j$ for $j\in S_{\mathrm{CEA}}$ and $\alpha$ for $j\in S_{\mathrm{CEA}}^c$
are therefore linear in $c_i$.

Continuity for $C>E$ holds by
\citet[Proposition~2]{moreno2006tal}.
\end{proof}

In the formulas below, use the sets and case conditions just to the left of $c_i$ for $g_i^-$
and just to its right for $g_i^+$, wherever that side exists.

\begin{lemma}[One-sided gains of CEA]\label{lem:cea-gains}
The one-sided gains of claimant $i$ are
\begin{equation}\label{eq:cea-gains}
g_i^{\pm}=
\begin{cases}
1, & i\in S_{\mathrm{CEA}},\\
0, & i\in S_{\mathrm{CEA}}^c,
\end{cases}
\end{equation}
and any kink occurs when claimant $i$ changes from $S_{\mathrm{CEA}}$ to $S_{\mathrm{CEA}}^c$,
with $(g_i^-,g_i^+)=(1,0)$.
\end{lemma}
\begin{proof}
By \cref{lem:cea-sides,lem:cea-pl}, each one-sided gain is the slope on an interval with fixed sets.

\emph{Slopes.} For $i\in S_{\mathrm{CEA}}$, $\phi^{\text{CEA}}_i=c_i$ has slope $1$.
For $i\in S_{\mathrm{CEA}}^c$, the equation
$\sum_{j\in S_{\mathrm{CEA}}}c_j$\allowbreak${}+|S_{\mathrm{CEA}}^c|\alpha$\allowbreak${}=E$ does not depend on $c_i$,
so $\phi^{\text{CEA}}_i=\alpha$ has slope $0$.

\emph{Kinks.} The gain formula depends only on claimant $i$'s membership,
so changes in rivals' membership alone create no kink.
By the proof of \cref{lem:cea-sides}, $i$ changes sets at most once,
from $S_{\mathrm{CEA}}$ to $S_{\mathrm{CEA}}^c$.
The gain then falls from $1$ to $0$.
\end{proof}

\begin{theorem}[Truthful region of CEA]\label{thm:cea}
The CEA rule has $\lo=0$ and $\hi=1$.
Hence it is $(p^T,p^F)$-truthful if and only if $p^T=0$ and $p^F\ge1$.
It is untruthful under every nondecreasing dependence with $\dep(1)>0$.
For illustration, under linear dependence it is truthful only at $\lambda=0$.
\end{theorem}

\begin{proof}
By \cref{lem:cea-gains}, every one-sided gain is $0$ or $1$.
Both values occur at $\claims=(\varepsilon,E,\dots,E)$ with $0<\varepsilon<E/N$.
Here $\alpha=\tfrac{E-\varepsilon}{N-1}$ satisfies $\varepsilon<\alpha<E$,
so claimant $1$ has gain $1$ and every rival has gain $0$.
Thus $\lo=0$ and $\hi=1$.

\emph{Dependence.} The truthfulness claims follow from \cref{prop:criterion,prop:robust}.
\end{proof}

\subsection{The Constrained Equal Losses Rule}\label{sec:cel}
Denote by $S_{\mathrm{CEL}}$ the claimants receiving $0$ under CEL, and by $S_{\mathrm{CEL}}^c$ its complement:
\begin{align*}
    S_{\mathrm{CEL}} &= \{j: \phi^{CEL}_j = 0\}
    \\
    S_{\mathrm{CEL}}^c &= \{j: \phi^{CEL}_j >0\}.
\end{align*}

\begin{lemma}[Finitely many changes of side under CEL]\label{lem:cel-sides}
Fix $E$ and $c_{-i}$.
As $c_i$ varies,
the sets $S_{\mathrm{CEL}}$ and $S_{\mathrm{CEL}}^c$ change at finitely many claims.
\end{lemma}
\begin{proof}
For $C\le E$, the sets are constant.

For $C>E$,
If $\beta$ decreased as $c_i$ increased, no award $\max(c_j-\beta,0)$ would decrease
and every positive award would increase.
Since $E>0$, at least one award is positive, so this would contradict efficiency.
Hence $\beta$ is nondecreasing in $c_i$.

Claimant $i$ belongs to $S_{\mathrm{CEL}}^c$ exactly when his award is positive.
This award is nondecreasing in $c_i$ by monotonicity
\citep[Proposition~4]{moreno2006tal},
so once positive it stays positive.
Thus $i$ can change only from $S_{\mathrm{CEL}}$ to $S_{\mathrm{CEL}}^c$.

Each rival's claim is fixed, so $c_j-\beta$ is nonincreasing.
Once nonpositive, it stays nonpositive, so a rival can change only from $S_{\mathrm{CEL}}^c$ to $S_{\mathrm{CEL}}$.
Each claimant therefore changes sets at most once, proving finiteness.
\end{proof}

\begin{lemma}[CEL is piecewise linear and continuous]\label{lem:cel-pl}
Fix $E$ and $c_{-i}$.
The award $\phi^{\text{CEL}}_i$ is continuous and piecewise linear in $c_i$.
\end{lemma}
\begin{proof}
By \cref{lem:cel-sides}, the sets are fixed between finitely many claims.
On each such interval with $C>E$, $S_{\mathrm{CEL}}^c$ is nonempty because $E>0$, and efficiency gives
\[
\beta=\frac{\sum_{j\in S_{\mathrm{CEL}}^c}c_j-E}{|S_{\mathrm{CEL}}^c|}.
\]
This expression is linear in $c_i$, as are the awards $0$
for $j\in S_{\mathrm{CEL}}$ and $c_j-\beta$ for $j\in S_{\mathrm{CEL}}^c$.

Continuity for $C>E$ holds by
\citet[Proposition~2]{moreno2006tal}.
\end{proof}

For both CEA and CEL, full payment adds at most one linear interval, on which $\phi_i=c_i$.
For $C>E$, feasibility and efficiency give $0\le c_i-\phi_i\le C-E$,
so this extension is continuous at $C=E$.

\begin{lemma}[One-sided gains of CEL]\label{lem:cel-gains}
The one-sided gains of claimant $i$ at $c_i\in\mathcal D_i(E,c_{-i})$ are
\begin{equation}\label{eq:cel-gains}
g_i^{\pm}=
\begin{cases}
0, & i\in S_{\mathrm{CEL}},\\
1-\tfrac1{|S_{\mathrm{CEL}}^c|}, & i\in S_{\mathrm{CEL}}^c,
\end{cases}
\qquad |S_{\mathrm{CEL}}^c|\ge2 .
\end{equation}
As $c_i$ increases, claimant $i$ enters $S_{\mathrm{CEL}}^c$ at $c_i=\beta$,
where $(g_i^-,g_i^+)=(0,1-\tfrac1{|S_{\mathrm{CEL}}^c|})$.
At all other kinks, the gain falls when rivals leave $S_{\mathrm{CEL}}^c$.
Rivals with equal claims leave together.
\end{lemma}
\begin{proof}
By \cref{lem:cel-sides,lem:cel-pl}, each one-sided gain is the slope on an interval with fixed sets.

\emph{Size of $S_{\mathrm{CEL}}^c$.} Since $\beta=0$ would give total award $C>E$, necessarily $\beta>0$.
Every positive award satisfies $c_j-\beta<c_j\le E$.
Efficiency then requires at least two positive awards, so $|S_{\mathrm{CEL}}^c|\ge2$.

\emph{Slopes.} For $i\in S_{\mathrm{CEL}}$, $\phi^{\text{CEL}}_i=0$ has slope $0$. For $i\in S_{\mathrm{CEL}}^c$, the defining
equation gives
\[
\beta=\frac{\sum_{j\in S_{\mathrm{CEL}}^c}c_j-E}{|S_{\mathrm{CEL}}^c|},
\]
so $\phi^{\text{CEL}}_i=c_i-\beta$ has slope $1-\tfrac1{|S_{\mathrm{CEL}}^c|}$.

\emph{Kinks.} By the proof of \cref{lem:cel-sides}, as $c_i$ increases,
claimant $i$ can only enter $S_{\mathrm{CEL}}^c$, and rivals can only leave.
When $i$ enters at $c_i=\beta$, \cref{eq:cel-gains} shows that the gain rises from $0$
to a positive value.

While $i\in S_{\mathrm{CEL}}$, the defining equation for $\beta$ does not depend on $c_i$,
so $\beta$ is constant and no rival leaves $S_{\mathrm{CEL}}^c$.
While $i\in S_{\mathrm{CEL}}^c$, rivals leave when $\beta$ reaches their claims,
so rivals with equal claims leave together.
Each such change reduces $|S_{\mathrm{CEL}}^c|$
and hence the gain $1-\tfrac1{|S_{\mathrm{CEL}}^c|}$.
\end{proof}

\begin{figure}[H]
\centering
\begin{tikzpicture}[scale=0.9]
\draw[->,thick] (0,0) -- (4,0) node[right] {$c_i$};
\draw[->,thick] (0,0) -- (0,3) node[above] {$\phi_i$};
\draw[ultra thick,blue] (0,0) -- (2,2) -- (3.7,2);
\draw[dashed] (2,2) -- (2,0) node[below] {$\alpha$};
\draw[dashed] (2,2) -- (0,2) node[left] {$\alpha$};
\filldraw (2,2) circle (1.6pt);
\node[blue] at (1.2,1.58) {\footnotesize slope $1$};
\node[blue] at (3.0,2.26) {\footnotesize slope $0$};
\node at (1.85,-0.9) {\small (a) CEA};
\begin{scope}[xshift=6cm]
\draw[->,thick] (0,0) -- (4.5,0) node[right] {$c_i$};
\draw[->,thick] (0,0) -- (0,3) node[above] {$\phi_i$};
\draw[ultra thick,red] (0,0) -- (1.3,0) -- (2.7,0.933) -- (4.2,1.683);
\draw (1.3,0.07) -- (1.3,-0.07) node[below] {$\beta$};
\filldraw (1.3,0) circle (1.6pt);
\filldraw (2.7,0.933) circle (1.6pt);
\draw[dashed] (2.7,0.933) -- (2.7,0) node[below] {\footnotesize $c_j=\beta$};
\node[red] at (0.65,0.26) {\footnotesize slope $0$};
\node[red] at (1.45,1.05) {\footnotesize slope $1-\tfrac1{|S_{\mathrm{CEL}}^c|}$};
\node[red] at (3.3,1.95) {\footnotesize slope $1-\tfrac1{|S_{\mathrm{CEL}}^c|}$, $|S_{\mathrm{CEL}}^c|$ smaller};
\node at (2.1,-0.9) {\small (b) CEL};
\end{scope}
\end{tikzpicture}
\caption{Kink geometry in claimant $i$'s own claim $c_i$, the other claims fixed
(\Cref{lem:cea-gains,lem:cel-gains}).
\textbf{(a)} CEA: the slope falls from $1$ to $0$ at $c_i=\alpha$.
\textbf{(b)} CEL: the slope rises at $c_i=\beta$ and falls when rivals leave
$S_{\mathrm{CEL}}^c$ as $c_i$ increases.}
\label{fig:kinks}
\end{figure}
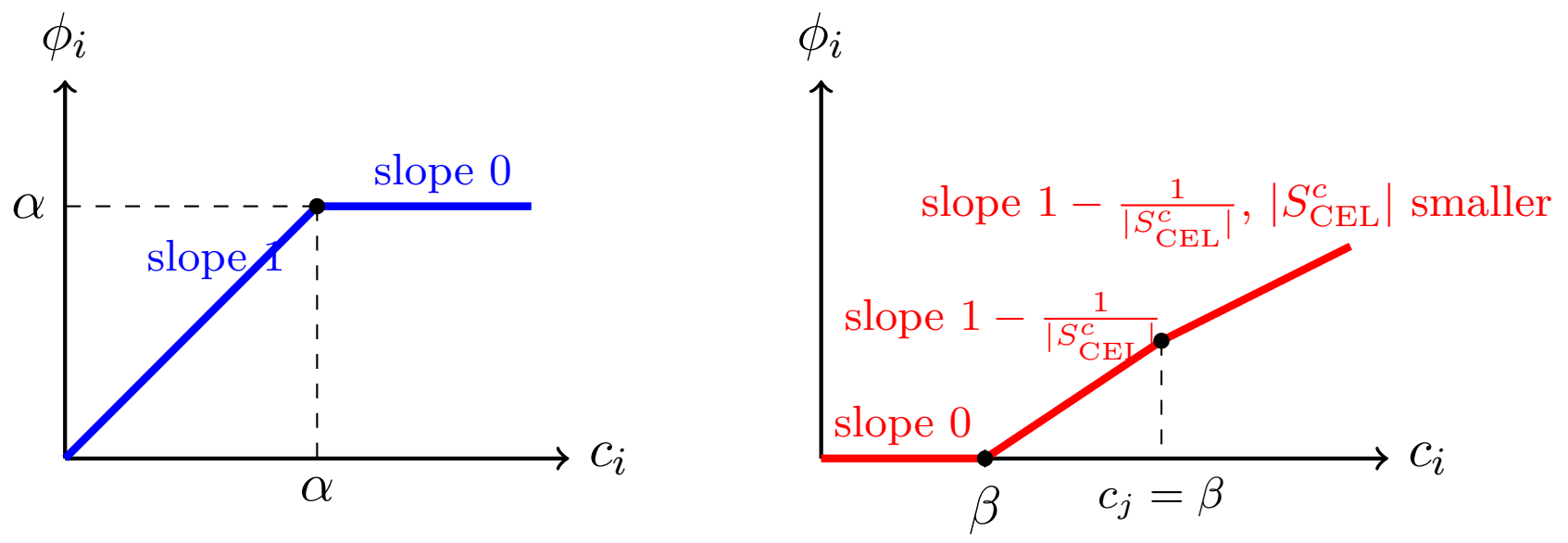

\Cref{fig:kinks} illustrates these changes for CEA and CEL.

\begin{theorem}[Truthful region of CEL for $N\ge3$]\label{thm:cel}
For $N\ge3$, the CEL rule has
$\lo=0$ and $\hi=1-\tfrac{1}{N}$.
Hence it is
$(p^T,p^F)$-truthful if and only if $p^T=0$ and $p^F\ge1-\tfrac{1}{N}$.
It is untruthful under every nondecreasing dependence with $\dep(1-\tfrac1N)>0$.
For illustration, under linear dependence it is truthful only at $\lambda=0$.
\end{theorem}

\begin{proof}
By \cref{lem:cel-gains}, every one-sided gain satisfies $0\le g_i^\pm\le1-\tfrac1N$.
Every claimant has gain $1-\tfrac1N$ when $c_j=c\in(E/N,E)$ for every $j$,
since $S_{\mathrm{CEL}}^c=[N]$.
Claimant $1$ has gain $0$ at $\claims=(\varepsilon,E,\dots,E)$ with
$0<\varepsilon<\beta=\tfrac{N-2}{N-1}E$, since claimant $1$ lies in $S_{\mathrm{CEL}}$.
Such an $\varepsilon$ exists because $N\ge3$.
Thus $\lo=0$ and $\hi=1-\tfrac1N$.

\emph{Dependence.} The truthfulness claims follow from \cref{prop:criterion,prop:robust}.
\end{proof}

\paragraph{Remark: CEL for $N=2$.}
The equality with CG holds under the model restriction $c_i,c_j\le E$.
For $c_i\in\mathcal D_i(E,c_{-i})$, each positive CEL award is strictly below its claim and hence below $E$.
Efficiency therefore requires both claimants to receive a positive award, so
$\beta=\tfrac{C-E}{2}$. With $j$ the rival claimant,
\[
\phi^{\text{CEL}}_i \;=\; c_i-\frac{C-E}{2} \;=\; \frac{E+c_i-c_j}{2} \;=\; \phi^{\text{CG}}_i ,
\]
where the last equality uses $\min(c_i,E)=c_i$ and $\min(c_j,E)=c_j$ in the CG formula.
Both rules also pay claims in full when $C\le E$, so they agree throughout the model domain.
They need not agree if a claim exceeds $E$: at $E=1$ and $(c_i,c_j)=(2,1)$, CEL awards
$(1,0)$ while CG awards $(\tfrac12,\tfrac12)$.
Thus CEL inherits $\lo=\hi=\tfrac12$ from \cref{thm:cg} below.

\subsection{The Contested Garment Rule}\label{sec:cg}

\begin{theorem}[Contested Garment rule]\label{thm:cg}
The CG rule for $N=2$ claimants has $\lo=\hi=\tfrac12$.
Hence it is $(p^T,p^F)$-truthful if and only if
$p^T\le\tfrac12\le p^F$.

For illustration, under linear dependence CG is truthful for every $\lambda\in[0,1]$:
$p^F\ge\tfrac12$ when $\lambda=0$, and
$p^F\in\bigl[\tfrac12,\tfrac{1}{2\lambda}\bigr]$ when $\lambda>0$.
Here $\lambda^{*}_{\text{CG}}=1$.
\end{theorem}

\begin{proof}
With both claims at most $E$, $\min(c_i,E)=c_i$ and $\min(c_j,E)=c_j$, so for $C>E$
\[
\phi^{\text{CG}}_i(E,\claims)=\tfrac12 c_i+\tfrac12(E-c_j).
\]
The marginal gain therefore equals $\tfrac12$
for every $c_i\in\mathcal D_i(E,c_{-i})$, and $\lo=\hi=\tfrac12$.
\Cref{prop:criterion} yields the region $p^T\le\tfrac12\le p^F$;
\Cref{cor:master} gives $\lambda^{*}_{\text{CG}}=1$ and the stated values of $p^F$.
\end{proof}

\begin{figure}[H]
\centering
\begin{tikzpicture}
 \draw[->, thick] (0,0) -- (0,6) node[above] {$p^F$};
 \draw[->, thick] (0,0) -- (3,0) node[right] {$p^T$};
 \coordinate (O) at (0,0);
 \fill[green!10] (0,1.5) rectangle (1.5,5.7);
 \draw[green!55!black] (1.5,1.5) -- (1.5,5.7);
 \draw[green!55!black] (0,1.5) -- (1.5,1.5);
 \node[right,green!45!black,xshift=8mm]
 at (0.8,4.5) {\footnotesize CG ($N=2$)};
 \node[below, green!45!black] at (1.5,0) {$\tfrac12$};
 \node[left,green!45!black,yshift=-1mm]
 at (0,1.5) {$\tfrac12$};
 \coordinate (Pft1) at (0,3);
 \coordinate (PftN) at (0,2);
 \draw[ultra thick, blue, ->] (Pft1) -- (0,5.8);
 \draw[very thick, red, dashed, ->] (PftN) -- (0,5.0);
 \filldraw (Pft1) circle (1.8pt) node[left] {$1$};
 \filldraw (PftN) circle (1.8pt)
 node[left,yshift=0.8mm] {$1-\tfrac1N$};
 \node[blue, left] at (0,5.5) {\footnotesize Prop, CEA};
 \node[red, left] at (0,4.5) {\footnotesize CEL ($N\ge3$)};
 \node[below] at (O) {$0$};
\end{tikzpicture}
\caption{The proportional rule and CEA require $p^T=0$ for truthfulness,
as does CEL for $N\ge3$ (\Cref{thm:prop,thm:cea,thm:cel}).
Among the displayed rules, only CG admits truthful pairs with $p^T>0$ (\Cref{thm:cg}).}
\label{fig:cel}
\end{figure}
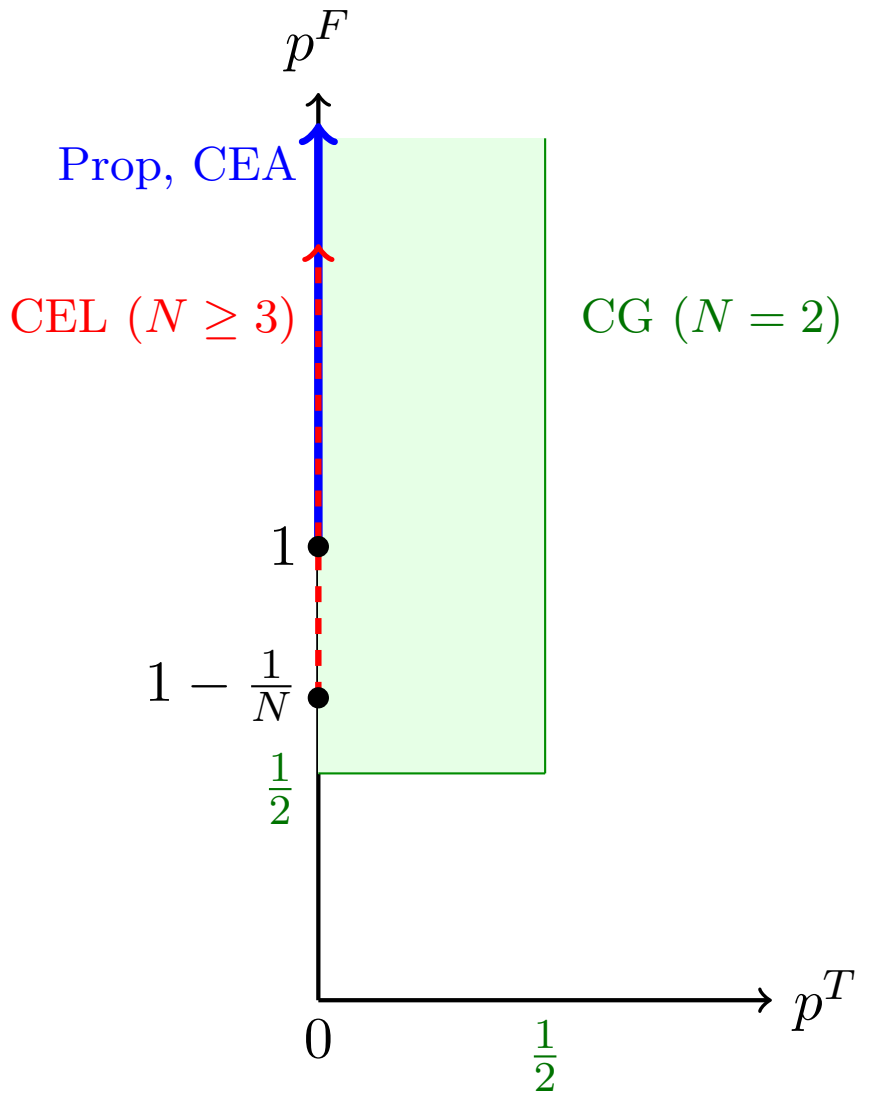

\Cref{fig:cel} shows the truthful regions of the classical rules.

\subsection{Constant Gains and the Two-Claimant Characterization}\label{sec:constant}

\begin{proposition}[Constant gain is impossible for $N\ge3$]\label{prop:impossible}
For $N\ge3$, no piecewise smooth rule has the same constant marginal gain $\mu$
for every claimant $i$ and every claim $c_i\in\mathcal D_i(E,c_{-i})$.
\end{proposition}
\begin{proof}
A common constant gain would give $\lo=\hi=\mu$.
For $N\ge3$, \cref{lem:dispersion} would then imply
\[
\mu=\lo\le\tfrac1N<\tfrac{N-1}{N}\le\hi=\mu,
\]
a contradiction.
\end{proof}

\begin{samepage}
\begin{theorem}[Rule with constant gain for two claimants]\label{thm:cg-constant}
Let $\phi$ be a piecewise smooth rule for two claimants.
If both claimants have the same constant marginal gain $\mu$
at every claim $c_i\in\mathcal D_i(E,c_{-i})$,
then $\phi$ equals the Contested Garment rule on the model domain $[0,E]^2$.
\end{theorem}
\end{samepage}

\begin{proof}
By \cref{lem:dispersion}, $\mu=\tfrac12$.
Fix $0<c_2\le E$. At $c_1=E-c_2$, full payment gives $\phi_1=E-c_2$.
For $c_1\in\mathcal D_1(E,c_2)=(E-c_2,E]$, integrating the constant gain gives
\[
\phi_1(E,\claims)=(E-c_2)+\int_{E-c_2}^{c_1}\tfrac12\,dt
=\frac{E+c_1-c_2}{2}=\phi^{\text{CG}}_1(E,\claims).
\]
The last equality uses $c_1,c_2\le E$.
Efficiency then gives $\phi_2=E-\phi_1=\phi^{\text{CG}}_2$.
\end{proof}

\begin{corollary}[Optimality of the Contested Garment rule]\label{cor:cg-optimal}
Let $N=2$. The truthful region of every piecewise smooth rule is contained in that of the Contested
Garment rule.
\end{corollary}

\begin{proof}
\Cref{lem:dispersion} gives $\lo\le\tfrac12\le\hi$ for every piecewise smooth rule at $N=2$, so
\cref{prop:criterion} places its truthful region inside that of the Contested Garment rule
(\cref{thm:cg}):
\[
\{(p^T,p^F):p^T\le\lo,\ p^F\ge\hi\}\;\subseteq\;\{(p^T,p^F):p^T\le\tfrac12\le p^F\},
\]
with equality forcing $\lo=\hi=\tfrac12$.
For every claimant, at each interior point
$c_i\in\mathcal D_i(E,c_{-i})$ where the award is differentiable,
$\tfrac12=\lo\le g_i^-=g_i^+\le\hi=\tfrac12$.
By piecewise smoothness, the remaining one-sided gains are limits of these derivatives and also equal $\tfrac12$.
Thus \cref{thm:cg-constant} identifies the rule as the Contested Garment rule.
\end{proof}

\Cref{prop:impossible} rules out a common constant marginal gain for $N\ge3$,
which explains the restriction to two claimants in \cref{thm:cg-constant}.

\subsection{The Parameterized TAL Rule and the Optimal $\theta$}\label{sec:tal}
For a parameter $\theta\in[0,1]$, denote by $S_{\mathrm{TAL}}$ the claimants receiving $\theta c_j$ under TAL-$\theta$,
and by $S_{\mathrm{TAL}}^c$ its complement:
\begin{align*}
S_{\mathrm{TAL}}&:=\{j:\phi_j^{\text{TAL-}\theta}=\theta c_j\},
\\
S_{\mathrm{TAL}}^c&:=[N]\setminus S_{\mathrm{TAL}}.
\end{align*}
For $C>E$, members of $S_{\mathrm{TAL}}^c$ receive $\alpha$ in the small estate domain $E\le\theta C$ and $c_j-\beta$ in the large estate domain $E>\theta C$.
At $E=\theta C$, the CEA branch gives $\phi_j^{\text{TAL-}\theta}\le\theta c_j$ for every $j$.
Since the awards sum to $E=\theta C$, all these inequalities are equalities.
Conversely, if every award equals $\theta c_j$, efficiency gives $E=\theta C$.
Thus $S_{\mathrm{TAL}}^c=\emptyset$ exactly at this boundary.
The one-sided gains there depend on the nearby sets, as detailed in \cref{lem:tal-kink-gains}.

\paragraph{The two parameters.}
For $C>E$ and $E\ne\theta C$, the set $S_{\mathrm{TAL}}^c$ is nonempty.
For $E<\theta C$, efficiency gives
\begin{equation}\label{eq:tal-alpha}
\alpha=\frac{E-\theta\sum_{j\in S_{\mathrm{TAL}}}c_j}{|S_{\mathrm{TAL}}^c|},
\end{equation}
and for $\theta C<E<C$ it gives
\begin{equation}\label{eq:tal-beta}
\beta=\frac{\sum_{j\in S_{\mathrm{TAL}}^c}c_j+\theta\sum_{j\in S_{\mathrm{TAL}}}c_j-E}{|S_{\mathrm{TAL}}^c|}
=\frac{\theta C+(1-\theta)\sum_{j\in S_{\mathrm{TAL}}^c}c_j-E}{|S_{\mathrm{TAL}}^c|}.
\end{equation}

\begin{lemma}[Finitely many changes of sets and of the sign of $E-\theta C$]\label{lem:tal-sides}
Fix $\theta\in[0,1]$, $E$, and $c_{-i}$.
As $c_i$ varies,
the sign of $E-\theta C$ and the sets $S_{\mathrm{TAL}}$, $S_{\mathrm{TAL}}^c$ change at finitely many claims.
\end{lemma}

\begin{proof}
\emph{Small estate domain.} Where $E<\theta C$, the rule is CEA with estate $E$ and claims $\theta\claims$.
$S_{\mathrm{TAL}}=S_{\mathrm{CEA}}$.
The estate and rival claims are fixed,
so \Cref{lem:cea-sides} implies finitely many changes of the sets.

\emph{Large estate domain.} When $\theta C<E<C$, each claimant receives $\theta c_j$ plus the CEL award with
estate $E-\theta C$ and claims $(1-\theta)\claims$.
These inputs vary linearly with $c_i$, and
\[
0<E-\theta C<(1-\theta)C
\]
because $E<C$.
$S_{\mathrm{TAL}}=S_{\mathrm{CEL}}$.
Since the CEL estate is positive, $S_{\mathrm{TAL}}^c$ is nonempty.
For each fixed nonempty $S_{\mathrm{TAL}}^c$, \cref{eq:tal-beta} makes $\beta$ linear in $c_i$.
The conditions $\beta\ge0$, $(1-\theta)c_j\le\beta$ for $j\in S_{\mathrm{TAL}}$, and
$(1-\theta)c_j>\beta$ for $j\in S_{\mathrm{TAL}}^c$
characterize that choice of sets.
These linear inequalities hold together on an interval, possibly empty or a single point.
Thus each fixed choice of sets can be entered and left at most once.
There are finitely many choices of $S_{\mathrm{TAL}}^c$, so the sets change only at finitely many interval endpoints.

\emph{Transition.} The expression $E-\theta(c_i+C_{-i})$ is linear and nonincreasing in $c_i$.
For $\theta>0$ it vanishes at most once, where $S_{\mathrm{TAL}}^c=\emptyset$; for $\theta=0$ it stays positive.
Thus there is at most one transition from CEL to CEA.

\emph{Full payment domain.}
For $C\le E$, the sets can change only at $c_i=0$ when $\theta<1$, and are constant when $\theta=1$.
The boundary $c_i=E-C_{-i}$ adds at most one point.
\end{proof}

\begin{lemma}[TAL--$\theta$ is piecewise linear and continuous]\label{lem:tal-cont}
Fix $\theta\in[0,1]$, $E$, and $c_{-i}$.
The award $\phi_i^{\text{TAL-}\theta}$ is continuous and piecewise linear in $c_i$.
\end{lemma}
\begin{proof}
For $C>E$, continuity holds by \citet[Proposition~2]{moreno2006tal}.

By \cref{lem:tal-sides}, the sets and the sign of $E-\theta C$ are fixed between finitely many points.
On each resulting interval with $C>E$,
\cref{eq:tal-alpha,eq:tal-beta} make the award linear in $c_i$.
Continuity joins these intervals, including at $E=\theta C$.

The full-payment claims satisfy $\phi_i^{\text{TAL-}\theta}=c_i$ and add at most one linear interval.
For $C>E$, feasibility and efficiency give $0\le c_i-\phi_i^{\text{TAL-}\theta}\le C-E$,
which proves continuity at $C=E$.
\end{proof}

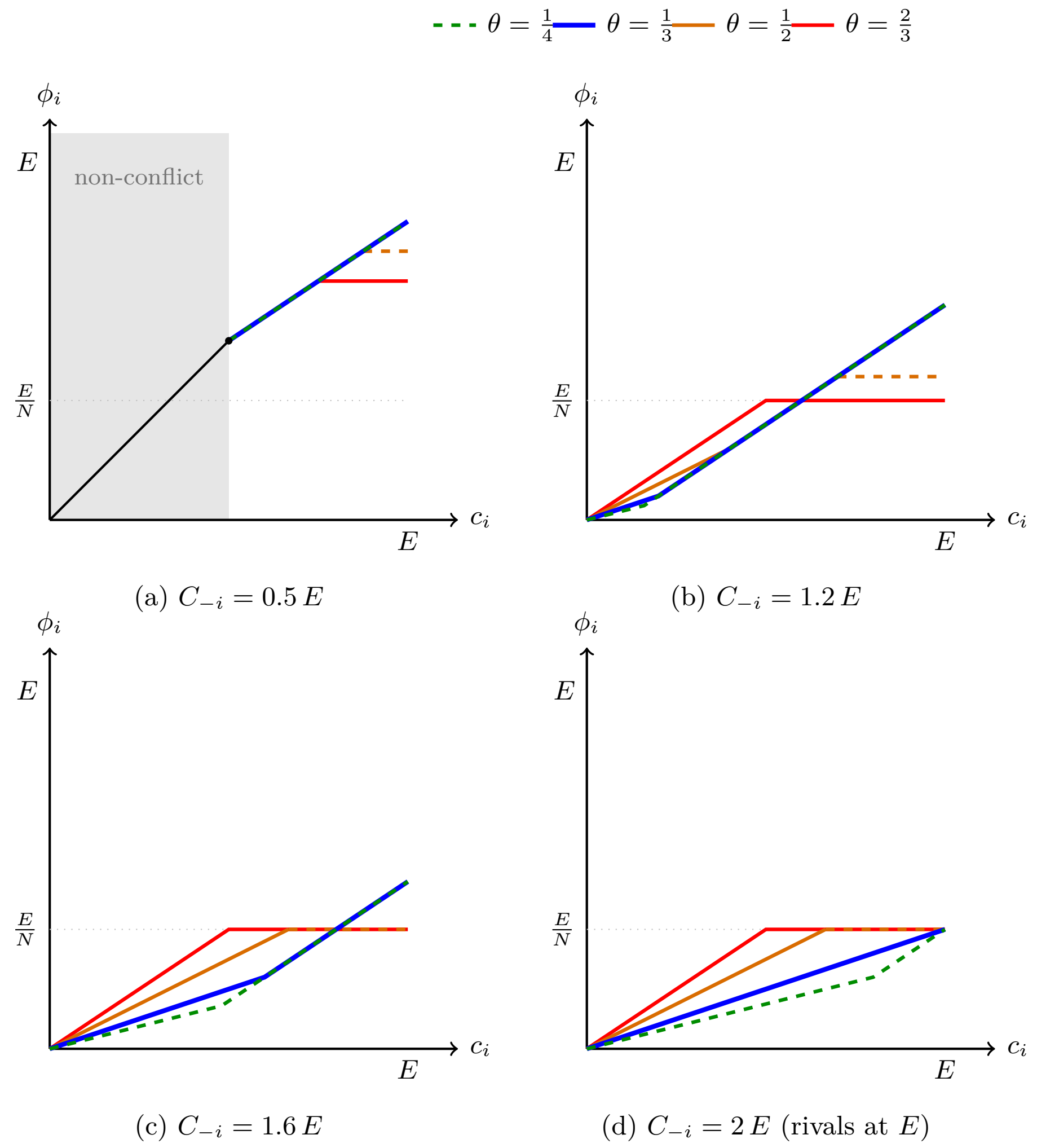
\begin{figure}[H]
\centering
\begin{tikzpicture}
\draw[very thick,green!55!black,dashed] (4.5,12.0) -- (5.0,12.0);
\node[right] at (5.0,12.0) {\footnotesize $\theta=\tfrac14$};
\draw[ultra thick,blue] (5.9,12.0) -- (6.4,12.0);
\node[right] at (6.4,12.0) {\footnotesize $\theta=\tfrac13$};
\draw[very thick,orange!85!black] (7.3,12.0) -- (7.8,12.0);
\node[right] at (7.8,12.0) {\footnotesize $\theta=\tfrac12$};
\draw[very thick,red] (8.7,12.0) -- (9.2,12.0);
\node[right] at (9.2,12.0) {\footnotesize $\theta=\tfrac23$};
\begin{scope}[xshift=0cm,yshift=6.2cm,xscale=4.2,yscale=4.2]
\fill[black!10] (0,0) rectangle (0.5,1.08);
\node[black!55] at (0.25,0.96) {\scriptsize non-conflict};
\draw[->,thick] (0,0) -- (1.14,0) node[right] {\footnotesize $c_i$};
\draw[->,thick] (0,0) -- (0,1.12) node[above] {\footnotesize $\phi_i$};
\node[below] at (1,0) {\footnotesize $E$}; \node[left] at (0,1) {\footnotesize $E$};
\node[left] at (0,0.3333) {\footnotesize $\tfrac{E}{N}$};
\draw[black!25,dotted] (0,0.3333) -- (1,0.3333);
\draw[thick,black] (0,0) -- (0.5,0.5); 
\draw[very thick,red] (0.5,0.5) -- (0.75,0.6667) -- (1,0.6667);
\draw[very thick,orange!85!black] (0.5,0.5) -- (0.875,0.75);
\draw[very thick,orange!85!black,dashed] (0.875,0.75) -- (1,0.75);
\draw[ultra thick,blue] (0.5,0.5) -- (1,0.8333);
\draw[very thick,green!55!black,dashed] (0.5,0.5) -- (1,0.8333); 
\node[fill=black,circle,inner sep=0pt,minimum size=2.5pt] at (0.5,0.5) {};
\node at (0.5,-0.22) {\footnotesize (a) $C_{-i}=0.5\,E$};
\end{scope}
\begin{scope}[xshift=6.3cm,yshift=6.2cm,xscale=4.2,yscale=4.2]
\draw[->,thick] (0,0) -- (1.14,0) node[right] {\footnotesize $c_i$};
\draw[->,thick] (0,0) -- (0,1.12) node[above] {\footnotesize $\phi_i$};
\node[below] at (1,0) {\footnotesize $E$}; \node[left] at (0,1) {\footnotesize $E$};
\node[left] at (0,0.3333) {\footnotesize $\tfrac{E}{N}$};
\draw[black!25,dotted] (0,0.3333) -- (1,0.3333);
\draw[very thick,red] (0,0) -- (0.5,0.3333) -- (1,0.3333);
\draw[very thick,orange!85!black] (0,0) -- (0.4,0.2) -- (0.7,0.4);
\draw[very thick,orange!85!black,dashed] (0.7,0.4) -- (1,0.4);
\draw[ultra thick,blue] (0,0) -- (0.2,0.0667) -- (1,0.6);
\draw[very thick,green!55!black,dashed] (0,0) -- (0.16,0.04) -- (1,0.6); 
\node at (0.5,-0.22) {\footnotesize (b) $C_{-i}=1.2\,E$};
\end{scope}
\begin{scope}[xshift=0cm,yshift=0cm,xscale=4.2,yscale=4.2]
\draw[->,thick] (0,0) -- (1.14,0) node[right] {\footnotesize $c_i$};
\draw[->,thick] (0,0) -- (0,1.12) node[above] {\footnotesize $\phi_i$};
\node[below] at (1,0) {\footnotesize $E$}; \node[left] at (0,1) {\footnotesize $E$};
\node[left] at (0,0.3333) {\footnotesize $\tfrac{E}{N}$};
\draw[black!25,dotted] (0,0.3333) -- (1,0.3333);
\draw[very thick,red] (0,0) -- (0.5,0.3333) -- (1,0.3333);
\draw[very thick,orange!85!black] (0,0) -- (0.6667,0.3333);
\draw[very thick,orange!85!black,dashed] (0.6667,0.3333) -- (1,0.3333);
\draw[ultra thick,blue] (0,0) -- (0.6,0.2) -- (1,0.4667);
\draw[very thick,green!55!black,dashed] (0,0) -- (0.48,0.12) -- (1,0.4667); 
\node at (0.5,-0.22) {\footnotesize (c) $C_{-i}=1.6\,E$};
\end{scope}
\begin{scope}[xshift=6.3cm,yshift=0cm,xscale=4.2,yscale=4.2]
\draw[->,thick] (0,0) -- (1.14,0) node[right] {\footnotesize $c_i$};
\draw[->,thick] (0,0) -- (0,1.12) node[above] {\footnotesize $\phi_i$};
\node[below] at (1,0) {\footnotesize $E$}; \node[left] at (0,1) {\footnotesize $E$};
\node[left] at (0,0.3333) {\footnotesize $\tfrac{E}{N}$};
\draw[black!25,dotted] (0,0.3333) -- (1,0.3333);
\draw[very thick,red] (0,0) -- (0.5,0.3333) -- (1,0.3333);
\draw[very thick,orange!85!black] (0,0) -- (0.6667,0.3333);
\draw[very thick,orange!85!black,dashed] (0.6667,0.3333) -- (1,0.3333);
\draw[ultra thick,blue] (0,0) -- (1,0.3333);
\draw[very thick,green!55!black,dashed] (0,0) -- (0.8,0.2) -- (1,0.3333); 
\node at (0.5,-0.22) {\footnotesize (d) $C_{-i}=2\,E$ (rivals at $E$)};
\end{scope}
\end{tikzpicture}
\caption{Own-claim award $\phi_i^{\text{TAL-}\theta}(c_i)$ at $N=3$, with the two rival claims both fixed at $C_{-i}/2$,
for $\theta\in\{\tfrac14,\tfrac13,\tfrac12,\tfrac23\}$ and four rival totals $C_{-i}$
(\cref{lem:tal-cont}). Here $\tfrac1N=\tfrac13$. Panel (a) includes the full-payment claims
$c_i\le E-C_{-i}$ (grey band), whereas $c_i\in\mathcal D_i(E,c_{-i})$ throughout panels (b)--(d).
For $\theta>\tfrac1N$, the award eventually stops rising, so $\lo=0$
(\cref{lem:tal-gains-enum}) and only $\lambda=0$ allows truthfulness under linear dependence
(\cref{cor:master}). For the displayed values
$0<\theta\le\tfrac1N$, the gain remains positive (\cref{lem:elim,lem:artifact}).
The bound $\lambda^{*}(\theta)$ is maximized at $\theta=\tfrac1N$ (\cref{thm:tal-lambda}).}
\label{fig:tal-theta}
\end{figure}

\Cref{fig:tal-theta} compares the own-claim awards as $\theta$ varies.

\begin{lemma}[One-sided gains of TAL--$\theta$]\label{lem:tal-gains-enum}
For $\theta\in[0,1]$ and $c_i\in\mathcal D_i(E,c_{-i})$, the one-sided gains are
\begin{equation}\label{eq:tal-gains-enum}
g_i^{\pm}=
\begin{cases}
\theta, & i\in S_{\mathrm{TAL}},\\
0, & i\in S_{\mathrm{TAL}}^c,\ E<\theta C,\\
1-\tfrac1{|S_{\mathrm{TAL}}^c|}, & i\in S_{\mathrm{TAL}}^c,\ E>\theta C.
\end{cases}
\end{equation}
\end{lemma}
\begin{proof}
By \cref{lem:tal-sides,lem:tal-cont}, each one-sided gain is the slope on an interval with fixed sets
and a fixed sign of $E-\theta C$.
For $i\in S_{\mathrm{TAL}}$, the award $\theta c_i$ has slope $\theta$.
For $i\in S_{\mathrm{TAL}}^c$ in the small estate, \cref{eq:tal-alpha} omits $c_i$, giving slope $0$.
For $i\in S_{\mathrm{TAL}}^c$ in the large estate, \cref{eq:tal-beta} gives
$\partial\beta/\partial c_i=1/|S_{\mathrm{TAL}}^c|$, so $c_i-\beta$ has slope $1-1/|S_{\mathrm{TAL}}^c|$.
\end{proof}

For $\theta>0$, \eqref{eq:tal-gains-enum} gives two cases of zero gain:
$i\in S_{\mathrm{TAL}}^c$ in the small estate domain, and
$S_{\mathrm{TAL}}^c=\{i\}$ in the large estate domain.
\Cref{lem:elim,lem:artifact} exclude both configurations on intervals with $c_i<E$
when $\theta\le1/N$. These intervals also determine the left-hand gain at $c_i=E$.

\begin{lemma}\label{lem:elim}
If $\theta\le 1/N$, then every $\claims\in[0,E]^N$ satisfies
$\theta C\le E$, with equality only when $\theta=1/N$ and $c_i=E$ for all $i$.
\end{lemma}
\begin{proof}
Since $C\le NE$ and $\theta\le1/N$, $\theta C\le\theta NE\le E$.
Equality forces $\theta=1/N$ and $C=NE$, hence $c_j=E$ for every $j$.
\end{proof}

\begin{lemma}\label{lem:artifact}
If $\theta\le 1/N$ and $c_i\in\mathcal D_i(E,c_{-i})$, then
\[
S_{\mathrm{TAL}}^c=\{i\}\ \Longrightarrow\ \theta=\tfrac1N\ \text{ and }\ c_i=E .
\]
\end{lemma}
\begin{proof}
Suppose $S_{\mathrm{TAL}}^c=\{i\}$.
By \cref{lem:elim}, $\theta C\le E$.
At equality $S_{\mathrm{TAL}}^c=\emptyset$, so $E>\theta C$ and \cref{eq:tal-beta} gives
\[
\beta=c_i+\theta C_{-i}-E.
\]
Each rival lies in $S_{\mathrm{TAL}}$, so $(1-\theta)c_j\le\beta$.
Summing gives $(1-\theta)C_{-i}\le(N-1)\beta$.
Substituting for $\beta$ and using $\theta\le1/N$ and $c_i\le E$ yields
\[
0\le(1-N\theta)C_{-i}\le(N-1)(c_i-E)\le0.
\]
All inequalities are therefore equalities.
Since $C_{-i}=C-c_i>E-c_i\ge0$, this forces $\theta=1/N$ and $c_i=E$.
\end{proof}

\begin{figure}[H]
\centering
\begin{tikzpicture}[xscale=7.5,yscale=9]
\draw[->,thick] (0,0) -- (1.1,0) node[right] {\footnotesize $c_i$};
\draw[->,thick] (0,0) -- (0,0.52) node[above] {\footnotesize $\phi_i$};
\node[below] at (1,0) {\footnotesize $E$};
\node[left] at (0,0.3333) {\footnotesize $\tfrac{E}{3}$};
\draw[black!25,dotted] (0,0.3333) -- (1,0.3333);
\draw[ultra thick,blue] (0,0) -- (0.7,0.2333) -- (0.9,0.3667) -- (1,0.4167);
\node[fill=black,circle,inner sep=0pt,minimum size=2.5pt] at (0.7,0.2333) {};
\node[fill=black,circle,inner sep=0pt,minimum size=2.5pt] at (0.9,0.3667) {};
\draw[dashed] (0.7,0.2333) -- (0.7,0) node[below] {\footnotesize $0.7E$};
\draw[dashed] (0.9,0.3667) -- (0.9,0) node[below] {\footnotesize $0.9E$};
\node[blue] at (0.5,0.11) {\footnotesize slope $\tfrac13$};
\node[blue] at (0.8,0.21) {\footnotesize slope $\tfrac23$};
\node[blue,right] at (1.0,0.42) {\footnotesize slope $\tfrac12$};
\end{tikzpicture}
\caption{Own-claim award of TAL--$\tfrac13$ at $N=3$, with rival claims $0.9E$ and $0.8E$.
The slope rises from $\theta=\tfrac13$ to $1-\tfrac1{|S_{\mathrm{TAL}}^c|}=\tfrac23$
(\cref{lem:tal-gains-enum}) at $c_i=0.7E$, where $i$ enters $S_{\mathrm{TAL}}^c$, and falls
to $\tfrac12$ at $c_i=0.9E$, where the rival with claim $0.8E$ leaves $S_{\mathrm{TAL}}^c$ and
$|S_{\mathrm{TAL}}^c|$ drops from $3$ to $2$.}
\label{fig:tal-leave}
\end{figure}
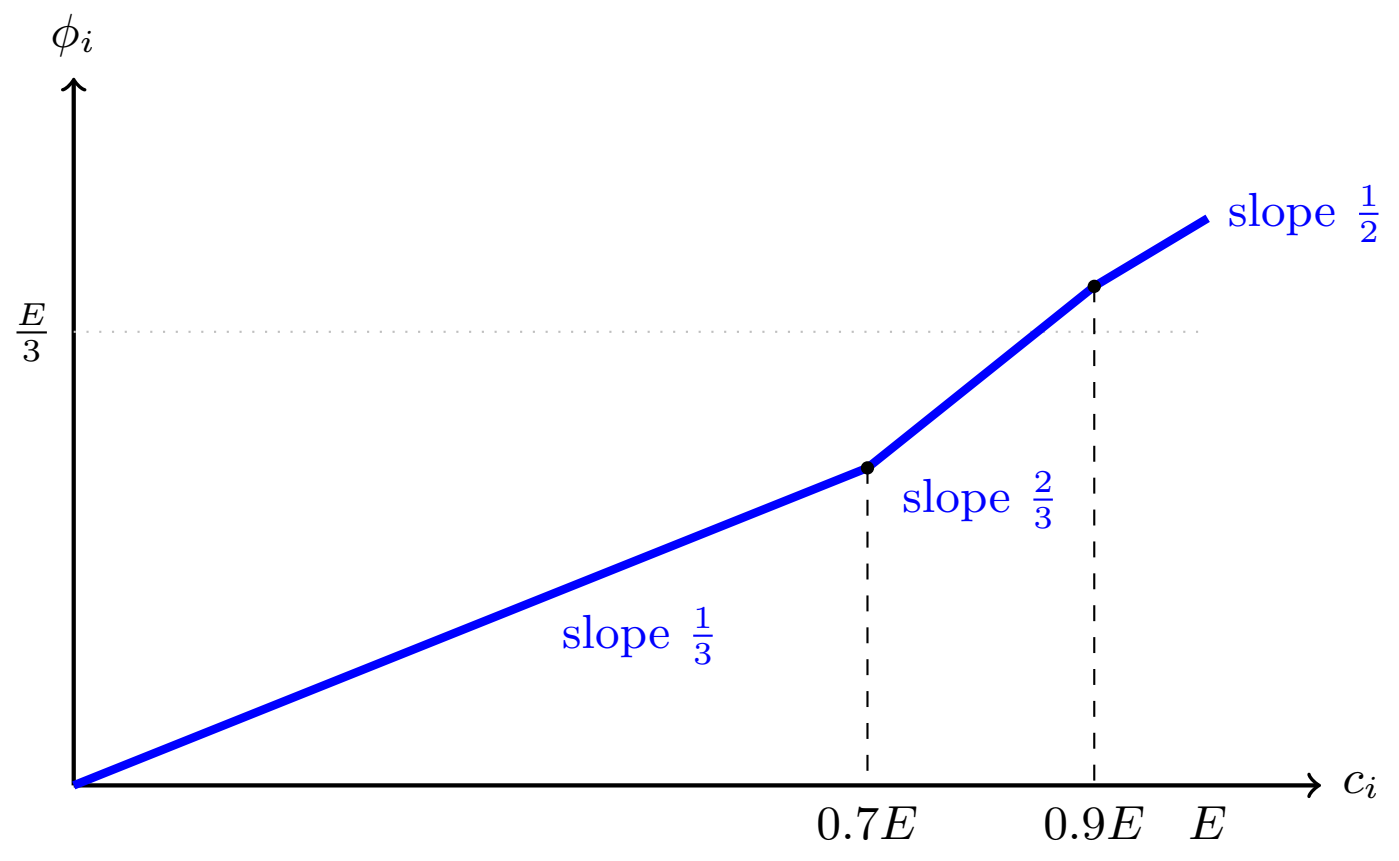

\Cref{fig:tal-leave} illustrates two of the kinks listed in \cref{app:tal-kinks}: the gain rises where
$i$ enters $S_{\mathrm{TAL}}^c$ and falls where a rival leaves it.

\begin{proposition}[Gain bounds for $\theta\le1/N$]\label{prop:gains}
Let $\theta\in[0,1/N]$. Each one-sided gain of $\phi_i^{\text{TAL-}\theta}$ at $c_i\in\mathcal D_i(E,c_{-i})$, wherever defined, lies
in
\[
\{\theta\}\cup\bigl\{1-\tfrac1k : k=2,\dots,N\bigr\}\subseteq\bigl[\theta,\,1-\tfrac1N\bigr].
\]
\end{proposition}
\begin{proof}
By \cref{lem:tal-sides,lem:tal-cont}, each one-sided gain is the slope on an interval
with $c_i<E$, including the left-hand slope at $c_i=E$.
On that interval, $\theta C<E$ by \cref{lem:elim}.
If $i\in S_{\mathrm{TAL}}$, \cref{lem:tal-gains-enum} gives gain $\theta$.
Otherwise, $i\in S_{\mathrm{TAL}}^c$ and \cref{lem:artifact} gives $|S_{\mathrm{TAL}}^c|\ge2$,
so \cref{lem:tal-gains-enum} gives gain $1-1/|S_{\mathrm{TAL}}^c|$.
Since $\theta\le\tfrac1N\le\tfrac12\le1-\tfrac1k\le1-\tfrac1N$ for $k=2,\dots,N$,
this proves both inclusions.
\end{proof}

\begin{theorem}[Truthful range of $\lambda$ for the TAL--$\theta$ family]\label{thm:tal-lambda}
For illustration, consider the linear dependence.
Within the TAL--$\theta$ family, truthfulness is possible exactly for
\[
\lambda\in
\begin{cases}
[0,1], & N=2,\ 0\le\theta\le \tfrac12,\\[1mm]
\bigl[0,\dfrac{\theta}{\,1-1/N\,}\bigr], & N\ge3,\ 0\le\theta\le \tfrac1N,\\[1mm]
\{0\}, & \tfrac1N<\theta\le1.
\end{cases}
\]
\end{theorem}

The proof is given in \cref{app:tal-lambda-proof}.

\begin{theorem}[Optimality of TAL--$1/N$]\label{thm:taln}
TAL--$1/N$ is the unique allocation rule in the TAL family that belongs to $\mathrm{OPT}_N$
(its truthful region is $0\le p^T\le\tfrac1N$, $p^F\ge1-\tfrac1N$).
Its truthful region contains that of every piecewise smooth allocation rule for $N$ claimants.
For $N\ge3$, the unique corresponding parameter is $\theta=1/N$.
For $N=2$, every $\theta\in[0,\tfrac12]$ defines this same rule, namely CG.
\end{theorem}
\begin{proof}
\emph{Optimality.} At $\theta=1/N$, \cref{prop:gains} gives $\lo\ge\tfrac1N$ and
$\hi\le1-\tfrac1N$; \cref{lem:dispersion} gives the reverse inequalities.
Thus $\lo=\tfrac1N$ and $\hi=1-\tfrac1N$.
By \cref{def:optimal} and its characterization \cref{eq:optimal-gains},
the rule belongs to $\mathrm{OPT}_N$.
Its truthful region is \cref{eq:optimal-region},
which contains every rule's truthful region by \cref{cor:pt-bound}.

\emph{Uniqueness within TAL.} Suppose TAL--$\theta$ belongs to $\mathrm{OPT}_N$.
Its truthfulness at $(1/N,1-1/N)$ implies that $\lambda=1/(N-1)$ is in its truthful range.
By \cref{thm:tal-lambda}, this forces $\theta=1/N$ for $N\ge3$ and $\theta\le\tfrac12$ for $N=2$.
In the latter case, all these parameters define CG, as shown in \cref{app:tal-lambda-proof}.
\end{proof}

It remains open whether any other allocation rule is in $OPT_N$ for some $N\geq 3$.

\newpage
\section{Strict Truthfulness}\label{sec:strict}

This section determines when truthful reporting is the unique best response.

\begin{definition}[Strict truthfulness]\label{def:strict}
Fix cost rates $(p^T,p^F)$.
A rule $\phi$ is \emph{strictly $(p^T,p^F)$-truthful} if,
for every $E>0$, $i\in[N]$, $c_{-i}\in[0,E]^{N-1}$ and
$c_i^*\in\mathcal D_i(E,c_{-i})$,
truthful reporting is the unique best response:
$U_i(c_i^*)>U_i(c_i)$ for every $c_i\in[0,E]$ with $c_i\ne c_i^*$.
\end{definition}

The sufficient condition for strict truthfulness is stronger than the criterion for weak truthfulness of \Cref{prop:criterion}.

\begin{proposition}[Sufficient condition for strict truthfulness]\label{prop:strict-criterion}
Let $\phi$ be piecewise smooth.
If
\[
p^T<\lo(\phi)\qquad\text{and}\qquad p^F>\hi(\phi),
\]
then $\phi$ is strictly $(p^T,p^F)$-truthful.
\end{proposition}

\begin{proof}
Fix $E,i,c_{-i}$ and $c_i^*\in\mathcal D_i(E,c_{-i})$.

Consider first an under-report $c_i<c_i^*$, 
and suppose the cost rates are $(\lo(\phi),\hi(\phi))$.
Both truthful reporting and under-reporting contain only true claims, 
therefore the costs are $\lo(\phi)c_i^*$ for claiming $c_i^*$ and $\lo(\phi)c_i$ for claiming $c_i$.

Weak-truthfulness (\Cref{prop:criterion}) gives
\[
\phi_i(E,(c_i^*,c_{-i}))-\lo(\phi)c_i^* = U_i(c_i^*)
\ge U_i(c_i) = \phi_i(E,(c_i,c_{-i}))-\lo(\phi)c_i.
\]
Hence
\[
\phi_i(E,(c_i^*,c_{-i}))-\phi_i(E,(c_i,c_{-i}))
\ge\lo(\phi)\cdot(c_i^*-c_i).
\]
At $p^T < \lo(\phi)$,
\[
\begin{aligned}
U_i(c_i^*)-U_i(c_i)
&=\phi_i(E,(c_i^*,c_{-i}))-\phi_i(E,(c_i,c_{-i}))
  -p^T(c_i^*-c_i)\\
&\ge[\lo(\phi)-p^T](c_i^*-c_i)>0.
\end{aligned}
\]

Consider now an over-report $c_i>c_i^*$. 
The true part remains $c_i^*$,
while the false part is $c_i-c_i^*$.
At the same cost rates $(\lo(\phi),\hi(\phi))$, weak truthfulness gives
\[
\phi_i(E,(c_i^*,c_{-i}))-\lo(\phi)c_i^*
\ge \phi_i(E,(c_i,c_{-i}))-\lo(\phi)c_i^*-\hi(\phi)(c_i-c_i^*).
\]
Hence
\[
\phi_i(E,(c_i,c_{-i}))-\phi_i(E,(c_i^*,c_{-i}))
\le\hi(\phi)\cdot(c_i-c_i^*).
\]
At $p^F > \hi(\phi)$,
\[
\begin{aligned}
U_i(c_i^*)-U_i(c_i)
&=p^F(c_i-c_i^*)-
  \bigl[\phi_i(E,(c_i,c_{-i}))-\phi_i(E,(c_i^*,c_{-i}))\bigr]\\
&\ge[p^F-\hi(\phi)](c_i-c_i^*)>0.
\end{aligned}
\]
In both cases the utility difference is positive, which completes the proof.
\end{proof}

The condition of \Cref{prop:strict-criterion} is sufficient but not necessary. For example, the proportional rule has $\lo=0$ and $\hi=1$ (\Cref{thm:prop}) but it is strictly $(0,1)$-truthful, as the following proposition shows. 
\begin{proposition}[Proportional]\label{prop:strict-proportional}
The proportional rule is strictly $(p^T,p^F)$-truthful
if and only if $p^T=0$ and $p^F\ge1$.
\end{proposition}

\begin{proof}
\emph{Necessity.} Strict truthfulness implies weak truthfulness.
By \cref{thm:prop}, this requires $p^T=0$ and $p^F\ge1$.

\emph{Sufficiency.} Fix these cost rates, $E,i,c_{-i}$,
and $c_i^*\in\mathcal D_i(E,c_{-i})$.
As in the proof of 
\Cref{thm:prop}, at every interior point in the conflict domain $t$,
\[
\frac{d}{dt}\phi_i^{\text{PROP}}(E,(t,c_{-i}))
=\frac{E C_{-i}}{(t+C_{-i})^2}>0.
\]
The allocation is continuous and strictly increasing,
so any under-report $c_i<c_i^*$ reduces utility when $p^T=0$:
\[
U_i(c_i^*)-U_i(c_i)
=\phi_i^{\text{PROP}}(E,(c_i^*,c_{-i}))
-\phi_i^{\text{PROP}}(E,(c_i,c_{-i}))>0.
\]
For an over-report $c_i>c_i^*$, 
since $c_i^*$ is in the conflict domain, every $t\in[c_i^*,c_i]$ satisfies $t+C_{-i}>E$, so
\[
\frac{E C_{-i}}{(t+C_{-i})^2}
\le\frac{E}{t+C_{-i}}<1\le p^F.
\]
Hence,
\[
U_i(c_i^*)-U_i(c_i)
=\int_{c_i^*}^{c_i}
  \left(p^F-\frac{E C_{-i}}{(t+C_{-i})^2}\right)\,dt>0.
\]
In both cases the utility difference is positive, which shows that the rule is strictly truthful.
\end{proof}

\begin{proposition}[CEA and CEL]\label{prop:strict-cea-cel}
CEA is not strictly $(p^T,p^F)$-truthful at any cost pair.
For $N\ge3$, the same holds for CEL.
\end{proposition}

\begin{proof}
Fix $E>0$ and an agent $i\in N$.
Assume that every agent other than $i$ claims $E$.

For CEA, assume that $i$'s true entitlement is $t\in(E/N,E)$. 
With $\alpha=E/N$,  every claimant receives $E/N$, so
\[
\phi_i^{\text{CEA}}(E,(t,c_{-i}))=E/N
\qquad\text{for every }t\in(E/N,E).
\]

For CEL with $N\ge3$, set $\beta=(N-2)E/(N-1)>0$, and assume that $i$'s true entitlement is $t\in(0,\beta)$. Then each rival receives $E/(N-1)$ and claimant $i$ receives zero:
\[
\phi_i^{\text{CEL}}(E,(t,c_{-i}))=0
\qquad\text{for every }t\in(0,\beta).
\]
For any $c_i<c_i^*$ in the conflict domain, under-reporting leaves the allocation unchanged
and saves $p^T(c_i^*-c_i)$:
\[
U_i(c_i^*)-U_i(c_i)=-p^T(c_i^*-c_i)\le0.
\]

For both rules there is a domain in which a false report does not change the outcome. Hence,
neither rule is strictly truthful.
\end{proof}

\noindent\begin{minipage}{\linewidth}
\begin{proposition}[TAL--$1/N$]\label{prop:strict-taln}
TAL--$1/N$ is strictly $(p^T,p^F)$-truthful
if and only if $p^T<\tfrac1N$ and $p^F>1-\tfrac1N$.
\end{proposition}

\begin{proof}
\emph{Sufficiency.} By \cref{thm:taln},
the extremal gains are $1/N$ and $1-1/N$.
The strict cost bounds therefore suffice by \cref{prop:strict-criterion}.

\emph{Necessity.} First fix $c_i^*=E$ and fix every rival claim at $E$.
Truthful reporting gives $E/N$, whereas reporting $0$ gives $0$.
Hence,
\[
U_i(E)-U_i(0)=(1/N-p^T)E,
\]
so strict truthfulness requires $p^T<1/N$.

Next choose $c_i^*\in(E/N,E)$ and set every rival claim equal to $c_i^*$.
Truthful reporting gives $E/N$.
If claimant $i$ reports $E$, then $C/N<E<C$ and the TAL formula with
$\beta=(1-1/N)c_i^*$ gives each rival $c_i^*/N$ and claimant $i$ the remainder
$E-(1-1/N)c_i^*$.
Thus,
\[
U_i(c_i^*)-U_i(E)=\bigl[p^F-(1-1/N)\bigr](E-c_i^*),
\]
so strict truthfulness requires $p^F>1-1/N$.

\end{proof}
\end{minipage}

\begin{corollary}[CG and CEL for $N=2$]\label{prop:strict-cg-cel}
For $N=2$, each of CG and CEL is strictly $(p^T,p^F)$-truthful
if and only if $p^T<\tfrac12<p^F$.
\end{corollary}

\begin{proof}
For $N=2$, CG and CEL both coincide with TAL--$1/2$ on the model domain $[0,E]^2$
by \cref{thm:taln,sec:cel}, so the result follows from \cref{prop:strict-taln}.
\end{proof}

For illustration, substituting $p^T=\lambda p^F$ in
\cref{prop:strict-cg-cel,prop:strict-taln} shows that strictly truthful cost rates exist
exactly when $0\le\lambda<1$ for CG and CEL with $N=2$,
and when $0\le\lambda<1/(N-1)$ for TAL--$1/N$.

\newpage
\section{The Price of Truthfulness}
\label{sec:welfare}
This section concerns weak truthfulness.
When $p^T>0$, truthful claimants still incur the cost of proving their claims.
Social welfare at a profile of reports $\claims$ is the sum of the claimants' net utilities,
$W(\claims):=\sum_{i\in[N]}U_i(E,\claims)$.
On the conflict domain, efficiency gives
\[
W(\claims)=E-\sum_{i\in[N]}\cost_i(c_i;c_i^*).
\]
When all claimants report truthfully, write $W=W(\claims^*)$.
The assumption $C^*>E$ gives
\[
\begin{gathered}
W=E-p^T C^*=E-\dep(p^F)C^*.
\end{gathered}
\]

\begin{proposition}[Nonnegative welfare under truthfulness]\label{prop:welfare-nonnegative}
Let $\phi$ be a piecewise smooth allocation rule
If $\phi$ is $(p^T,p^F)$-truthful, then social welfare at truthful reporting satisfies $W\ge0$.
Equality holds if and only if $p^T=\tfrac1N$ and $c_i^*=E$ for every $i\in[N]$.
\end{proposition}

\begin{proof}
By \cref{cor:pt-bound}, truthfulness requires $p^T\le\tfrac1N$.
Since $c_i^*\le E$ for every $i$,
$C^*\le NE$, and hence
\[
\begin{gathered}
W=E-p^T C^*\ge E-\frac{C^*}{N}\ge0.
\end{gathered}
\]
Both inequalities are equalities exactly when $p^T=\tfrac1N$ and
$C^*=NE$.
The latter condition holds if and only if $c_i^*=E$ for every $i\in[N]$.
\end{proof}

\begin{proposition}[Upper bound on truthful welfare]\label{prop:welfare-upper-bound}
Let $\phi$ be a piecewise smooth allocation rule.
If $\phi$ is $(p^T,p^F)$-truthful with $p^T=\dep(p^F)$, then
\[
\begin{gathered}
W\le E-\dep\!\left(1-\frac1N\right)C^*.
\end{gathered}
\]
\end{proposition}

\begin{proof}
By \cref{cor:pt-bound}, truthfulness requires $p^F\ge1-1/N$.
Since $\dep$ is nondecreasing, $p^T=\dep(p^F)\ge\dep(1-1/N)$.
Substituting into
$W=E-p^T C^*$ gives the bound.
\end{proof}

For the remainder of this section, $\dep(1-1/N)\le1/N$, so truthfulness is feasible.

\begin{proposition}[Optimal rules maximize truthful welfare]\label{prop:welfare-optimal}
For any $N\geq 2$ claimants,
every rule $\phi\in\mathrm{OPT}_N$ maximizes $W$ under the condition of truthfulness.
\end{proposition}

\begin{proof}
Choose
\[
p^F=1-\frac1N,
\qquad p^T=\dep\!\left(1-\frac1N\right).
\]
For every $\phi\in\mathrm{OPT}_N$, these cost rates preserve truthfulness by
\cref{eq:optimal-region} and the assumption $\dep(1-1/N)\le1/N$.
They also satisfy $p^T\le1/N\le1-1/N=p^F$.
At these rates, welfare attains the upper bound in \cref{prop:welfare-upper-bound}, so
\[
\begin{gathered}
W_{\mathrm{opt}}=E-\dep\!\left(1-\frac1N\right)C^*.
\end{gathered}\qedhere
\]
\end{proof}

\begin{samepage}
\begin{corollary}[Welfare optimality of TAL--$1/N$]\label{cor:tal-welfare}

TAL--$1/N$ maximizes the social welfare $W$ subject to truthfulness.
\end{corollary}

\begin{proof}
By \cref{thm:taln}, TAL--$1/N$ belongs to $\mathrm{OPT}_N$.
The conclusion follows from \cref{prop:welfare-optimal}.
\end{proof}
\end{samepage}

Maximizing $W$ means minimizing $p^T$, ideally to zero.
Since $p^T=\dep(p^F)$, reducing $p^F$ can reduce this cost, but truthfulness requires $p^F\ge1-1/N$.
The objective is therefore to lower costs as far as truthfulness permits.
Every rule in $\mathrm{OPT}_N$ attains the maximum welfare with the cost rates chosen above,
but rules outside this class may attain it as well.
For example, if $\dep$ is identically zero, the proportional rule also attains $W=E$ at $p^F=1$
(\cref{thm:prop}).

\newpage
\section{Conclusion and Future Work}\label{sec:synthesis}

For a fixed number of claimants $N$, TAL--$1/N$ has the largest truthful cost region under set inclusion
among all piecewise smooth allocation rules (\cref{thm:taln}).
The class $\mathrm{OPT}_N$ consists of all rules achieving this region (\cref{def:optimal}).
For two claimants, CG is the unique rule in $\mathrm{OPT}_2$ (\cref{cor:cg-optimal}).
For $n$ claimants, 
TAL--$1/N$ is the unique rule in $\mathrm{OPT}_N$ (\cref{thm:taln})
within the TAL family; whether there are other optimal rules not from this family remains open.
~
\Cref{tab:summary} summarizes the results rule by rule.

\begin{table}[H]
\centering
\setlength{\tabcolsep}{4pt}
\renewcommand{\arraystretch}{1.2}
\resizebox{\textwidth}{!}{%
\begin{tabular}{lcccllc}
\hline
Rule & $\lo$ & $\hi$ & $\lambda^{*}=\lo/\hi$ & \shortstack{Independent\\costs} & \shortstack{\strut Linear dependence\\($\lambda\in[0,1]$)}
& $\phi\in\mathrm{OPT}_N$\\
\hline
Proportional & $0$ & $1$ & $0$ & \shortstack[l]{\strut strict at\\$p^T{=}0,p^F{\ge}1$}
& strict iff $\lambda=0$ & No\\
CEA & $0$ & $1$ & $0$ & weak only & weak iff $\lambda=0$ & No\\
CEL \ ($N\ge3$) & $0$ & $1-\tfrac1N$ & $0$ & weak
only & weak iff $\lambda=0$ & No\\
CG \ ($N=2$) & $\tfrac12$ & $\tfrac12$ & $1$ & strict & strict
iff $\lambda<1$ & \textbf{Yes}\\
$\text{TAL}$--$1/N$ & $\tfrac1N$ & $1-\tfrac1N$ &
$\tfrac{1}{N-1}$ & strict & strict iff $\lambda<\tfrac{1}{N-1}$ & \textbf{Yes}\\
\hline
\end{tabular}}
\caption{Extremal marginal gains, truthfulness, and optimality.
The last column records membership in $\mathrm{OPT}_N$, the class of rules with the largest
weakly truthful cost region for the same $N$ (\cref{def:optimal,cor:cg-optimal,thm:taln}).
The $\lambda^{*}$ and linear-dependence columns concern the illustrative relation $p^T=\lambda p^F$.
CG allows only weak truthfulness at $\lambda=1$.
$\text{TAL}$--$1/N$ allows only weak truthfulness at $\lambda=1/(N-1)$.
For $N=2$,
CEL (and every $\text{TAL}$--$\theta$ with $\theta\le\tfrac12$) equals CG
(\Cref{sec:cel,sec:tal}).}
\label{tab:summary}
\end{table}

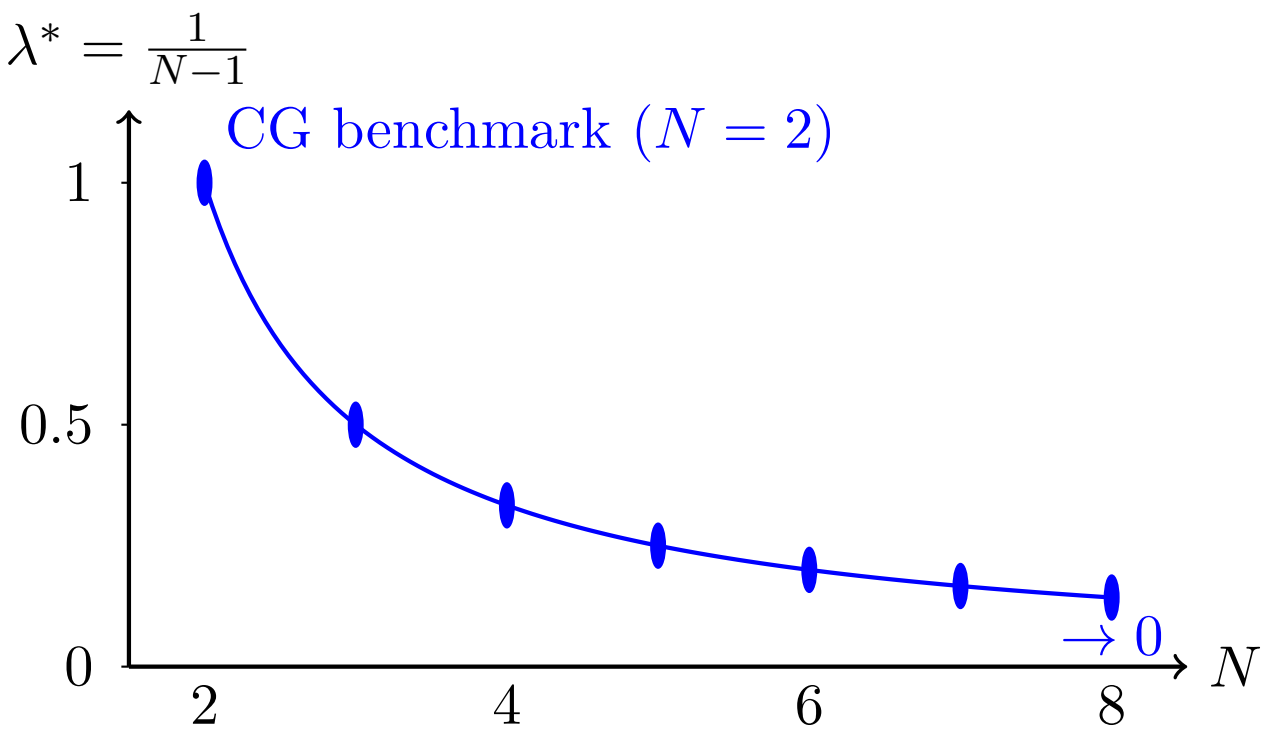
\begin{figure}[H]
\centering
\begin{tikzpicture}[xscale=1.0,yscale=3.2]
 \draw[->,thick] (1.5,0) -- (8.5,0) node[right] {$N$};
 \draw[->,thick] (1.5,0) -- (1.5,1.15) node[above] {$\lambda^{*}=\tfrac{1}{N-1}$};
 \foreach \y/\yl in {0/0,0.5/0.5,1/1}
   \draw (1.45,\y) -- (1.5,\y) node[left,xshift=-1mm] {\yl};
 \foreach \nn in {2,3,4,5,6,7,8}
   \filldraw[blue] (\nn,{1/(\nn-1)}) circle (1.3pt);
 \draw[blue,thick,domain=2:8,samples=120] plot (\x,{1/(\x-1)});
 \node[blue,above right] at (2,1) {CG benchmark ($N=2$)};
 \node[blue,below] at (8,{1/7}) {$\to 0$};
 \foreach \nn in {2,4,6,8} \node[below] at (\nn,0) {\nn};
\end{tikzpicture}
\caption{The bound $\lambda^{*}_{\text{TAL-}1/N}=1/(N-1)$ under the illustrative linear dependence,
against the number of claimants $N$.}
\label{fig:lambda}
\end{figure}

We see several opportunities for extending this work beyond the settings our model covers. 

The first concerns the range of admissible claims: we assume that no claimant can claim more than the entire estate ($c_i \le E$), which holds across a large space of instances, most naturally disputes over ownership of a fixed object. When claims arise from loans or other debts a single creditor may be owed more than the estate can ever pay, and extending the analysis there is substantive, since this restriction is what gives the Contested Garment rule its constant marginal gain. 

The second concerns the shape of litigation costs: we use the typical case in which they are linear in the claimed amounts, as with lawyers’ fees set as a percentage of the claim or penalties that rise with the amount fabricated. Fixed filing fees, convex costs arising when detection grows more likely as the fabrication grows, claimant-specific rates, and contingency fees tied to the award rather than the claim would each reshape the analysis in a different direction.  

\appendix
\newpage
\section{Kinks and One-Sided Gains}\label{app:tal-kinks}\label{app:enum}

Fix $E$ and $c_{-i}$. The cases below cover all kinks of the classical and TAL rules as $c_i$ varies.
Only interior claims can be kinks; at $c_i=0$ or $c_i=E$, only the inward gain is defined.
At $E=\theta C$, all awards equal $\theta c_j$, but the nearby sets determine the one-sided gains.
The change between the CEA and CEL formulas need not produce a kink in the award.

\begin{lemma}[One-sided gains at $E=\theta C$]\label{lem:tal-kink-gains}
For $c_i\in\mathcal D_i(E,c_{-i})\cap(0,E)$ and $E=\theta C$, the gain pair is
\begin{equation}\label{eq:tal-boundary-gains}
(g_i^-,g_i^+)=
\begin{cases}
(\theta,\theta), & c_i<\max_{j\ne i}c_j,\\[1mm]
\bigl(\min\{\theta,1-\tfrac1m\},0\bigr), & c_i\ge\max_{j\ne i}c_j,
\end{cases}
\end{equation}
where $m=|\{j\in[N]:c_j=c_i\}|$ in the second case.
There is a kink exactly when $i$ shares the largest claim with a rival.
If $i$ has the unique largest claim, both gains are zero.
At an endpoint, only the inward gain is defined.
\end{lemma}
\begin{proof}
Fix $E$ and $c_{-i}$. Since $\theta C=E<C$, $0<\theta<1$.
Efficiency gives
\[
\begin{aligned}
\sum_j\max\{0,(1-\theta)c_j-\beta\}&=E-\theta C && (E>\theta C),\\
\sum_j\max\{0,\theta c_j-\alpha\}&=\theta C-E && (E<\theta C).
\end{aligned}
\]
As $c_i$ approaches the boundary, these sums tend to zero.
Thus $\beta$ approaches $(1-\theta)\max_j c_j$ from the large estate,
and $\alpha$ approaches $\theta\max_j c_j$ from the small estate.
Only claimants with a largest claim at the boundary can belong to $S_{\mathrm{TAL}}^c$ nearby.
If $c_i<\max_{j\ne i}c_j$, then $i\in S_{\mathrm{TAL}}$ on both sides,
so $(g_i^-,g_i^+)=(\theta,\theta)$.

Now suppose $c_i\ge\max_{j\ne i}c_j$, and let $m=|\{j\in[N]:c_j=c_i\}|$
count the largest claims at the boundary.
For sufficiently small $\varepsilon>0$, the parameter at $c_i+\varepsilon$ is $\alpha=\theta c_i$,
so $S_{\mathrm{TAL}}^c=\{i\}$ and $g_i^+=0$.

At $c_i-\varepsilon$, every rival tied with $i$ at the boundary belongs to $S_{\mathrm{TAL}}^c$.
Including $i$ gives, by \cref{eq:tal-beta},
\[
\beta=(1-\theta)c_i-\frac{\varepsilon}{m},
\qquad
(1-\theta)(c_i-\varepsilon)>\beta\iff\theta>1-\frac1m.
\]
This choice of sets is valid exactly when the strict inequality holds, giving $g_i^-=1-1/m$.
Otherwise, $i\in S_{\mathrm{TAL}}$ and $g_i^-=\theta$; at equality, the two gain formulas agree.
Hence
\[
(g_i^-,g_i^+)=\bigl(\min\{\theta,1-\tfrac1m\},0\bigr)
\qquad\text{when }c_i\ge\max_{j\ne i}c_j.
\]
For $m=1$, both gains are zero; for $m\ge2$, the left gain is positive.
Thus a kink occurs exactly when $i$ shares the largest claim with a rival.
At an endpoint, only the inward gain is defined, so the estate boundary is not a kink there.
\end{proof}

For example, in panel (b) of \cref{fig:tal-theta}, $\theta=2/3$ and $c_i=3E/10$
give $E=\theta C$, but both rivals claim $3E/5$.
The gains are $(2/3,2/3)$, so there is no kink at this change of formula.
With three equal claims of $E/2$ and $\theta=2/3$, the boundary instead has gains $(2/3,0)$.

\paragraph{TAL away from $E=\theta C$.}
In the small estate domain $E<\theta C$, the only kink occurs when $i$ enters
$S_{\mathrm{TAL}}^c$ at $\theta c_i=\alpha$, with $(g_i^-,g_i^+)=(\theta,0)$.
Indeed, $\alpha$ is nonincreasing by the CEA argument in \cref{lem:cea-sides},
and \cref{lem:tal-gains-enum} shows that rivals changing sets alone do not change $i$'s gain.

\begin{samepage}
In the large estate domain $\theta C<E<C$, the possible gain changes are as follows.
Each cardinality is evaluated on the side indicated by its column.
\begin{center}
\begin{tabular}{@{}lcc@{}}
\hline
Change as $c_i$ increases & $g_i^-$ & $g_i^+$ \\
\hline
$i$ enters $S_{\mathrm{TAL}}^c$ & $\theta$ & $1-\tfrac1{|S_{\mathrm{TAL}}^c|}$ \\[1mm]
$i$ leaves $S_{\mathrm{TAL}}^c$ & $1-\tfrac1{|S_{\mathrm{TAL}}^c|}$ & $\theta$ \\[1mm]
Rivals leave; $i$ stays in $S_{\mathrm{TAL}}^c$ & $1-\tfrac1{|S_{\mathrm{TAL}}^c|}$ & $1-\tfrac1{|S_{\mathrm{TAL}}^c|}$ \\
\hline
\end{tabular}
\end{center}
\end{samepage}
Claimant $i$ changes sets at $(1-\theta)c_i=\beta$; a rival $j$ leaves at $(1-\theta)c_j=\beta$.
Rivals with equal claims leave together.
The first two rows also cover rivals leaving at the same claim where $i$ changes sets:
the cardinalities include all changes on the corresponding side.
There is a kink exactly when the two displayed gains differ.
If $i$ stays in $S_{\mathrm{TAL}}$, both gains are $\theta$, even if rivals leave.

These formulas follow from \cref{lem:tal-gains-enum}.
To see that the list is exhaustive, first observe that $\beta$ is nondecreasing in $c_i$.
If it decreased, no award $\max\{\theta c_j,c_j-\beta\}$ would decrease,
and every claimant initially in the nonempty set $S_{\mathrm{TAL}}^c$ would receive more,
contradicting efficiency.
Thus rivals can only leave $S_{\mathrm{TAL}}^c$.
By \cref{lem:tal-sides,lem:tal-cont}, a kink requires a change of sets or a change
between the large and small estate domains.
The table covers every change affecting $i$'s gain in the large estate;
the small estate case is covered above, and the domain boundary by \cref{lem:tal-kink-gains}.

\paragraph{The classical rules.}
The cases $\theta=0$ and $\theta=1$ give CEL and CEA, respectively.
Their kinks and proofs also appear in \cref{lem:cel-gains,lem:cea-gains}.
PROP and CG have no kinks within the conflict domain:
their own-claim awards are smooth there, as shown in the proofs of \cref{thm:prop,thm:cg}.

\paragraph{The full-payment boundary.}
For completeness, suppose $C=E$ at an interior claim $c_i\in(0,E)$.
Full payment gives $g_i^-=1$ for every rule.
For TAL with $\theta<1$, every positive claim remains in $S_{\mathrm{TAL}}^c$
when $c_i$ increases by a sufficiently small $\varepsilon>0$.
By \cref{eq:tal-beta}, $\beta=\varepsilon/|\{j:c_j>0\}|$, so
$g_i^+=1-1/|\{j:c_j>0\}|<1$; hence there is a kink.
For $\theta=1$, CEA has $g_i^+=1$ if $c_i<\max_{j\ne i}c_j$, and $g_i^+=0$ otherwise:
in the first case $i$ still receives his claim just above the boundary,
whereas in the second case the cap stays at the boundary value $c_i$.
Thus CEA has a kink here exactly when $i$ has a largest claim.
For PROP and CG, differentiating the formulas in \cref{thm:prop,thm:cg} gives
$g_i^+=1-c_i/E$ and $g_i^+=1/2$, respectively, so both have a kink here.
Below this boundary every rule pays $c_i$ and has no kink.

\newpage
\section{Proof of the TAL--$\theta$ Truthful Range}\label{app:tal-lambda-proof}

\begin{proof}[Proof of \Cref{thm:tal-lambda}]
By \cref{cor:master}, it suffices to compute $\lo$ and $\hi$.

If $N=2$ and $\theta\le\tfrac12$, then $\theta C\le E$ by \cref{lem:elim}.
For $\theta C<E<C$ and $j\ne i$, the CG award is $c_i-(C-E)/2$ and is at least
$\theta c_i$, since
\[
(1-\theta)c_i-\frac{C-E}{2}=\frac{(1-2\theta)c_i+E-c_j}{2}\ge0.
\]
The CG awards sum to $E$, so they satisfy the TAL--$\theta$ formula.
Continuity covers $E=\theta C$ (\cref{lem:tal-cont}), and \cref{thm:cg} gives
$\lo=\hi=\tfrac12$.

For $N\ge3$ and $\theta\le1/N$, \cref{prop:gains} bounds every gain between
$\theta$ and $1-\tfrac1N$.
At $c_i=0$ with every rival claim equal to $E$, the rivals receive
$E/(N-1)>\theta E$ by symmetry and efficiency, and $\beta=E-E/(N-1)>0$.
Thus $i\in S_{\mathrm{TAL}}$ for claims just above zero, giving gain $\theta$.

At $c_j=E/2$ for every $j$, $\theta C<E$, and symmetry and efficiency give each claimant
$E/N>\theta E/2$, so $S_{\mathrm{TAL}}^c=[N]$ and the gain is $1-\tfrac1N$.
The strict inequalities make both gains occur on intervals, so
$\lo=\theta$ and $\hi=1-\tfrac1N$.

If $\theta>1/N$, set every claim equal to $E$.
Then $\theta C>E$ and every scaled claim exceeds $E/N$.
These inequalities persist as $c_i$ decreases slightly, so the small-estate award
remains $E/N$ on an interval and $\lo=0$.

\Cref{cor:master} gives the stated ranges.
For $N\ge3$, their right endpoint increases strictly up to $\theta=1/N$ and is zero
for larger $\theta$; for $N=2$, every $\theta\le\tfrac12$ gives the full range $[0,1]$.
\end{proof}
\newpage

\bibliographystyle{plainnat}
\IfFileExists{artical/references.bib}{%
  \bibliography{artical/references}%
}{%
  \bibliography{references}%

\begin{thebibliography}{27}
\providecommand{\natexlab}[1]{#1}
\providecommand{\url}[1]{\texttt{#1}}
\expandafter\ifx\csname urlstyle\endcsname\relax
  \providecommand{\doi}[1]{doi: #1}\else
  \providecommand{\doi}{doi: \begingroup \urlstyle{rm}\Url}\fi

\bibitem[Ashlagi et~al.(2012)Ashlagi, Karag{\"o}zo{\u{g}}lu, and
  Klaus]{ashlagi2012estate}
I.~Ashlagi, E.~Karag{\"o}zo{\u{g}}lu, and B.~Klaus.
\newblock A non-cooperative support for equal division in estate division
  problems.
\newblock \emph{Mathematical Social Sciences}, 63\penalty0 (3):\penalty0
  228--233, 2012.

\bibitem[Aumann and Maschler(1985)]{aumann1985game}
R.~J. Aumann and M.~Maschler.
\newblock Game theoretic analysis of a bankruptcy problem from the {Talmud}.
\newblock \emph{Journal of Economic Theory}, 36\penalty0 (2):\penalty0
  195--213, 1985.

\bibitem[Ben-Porath et~al.(2014)Ben-Porath, Dekel, and
  Lipman]{benporath2014optimal}
E.~Ben-Porath, E.~Dekel, and B.~L. Lipman.
\newblock Optimal allocation with costly verification.
\newblock \emph{American Economic Review}, 104\penalty0 (12):\penalty0
  3779--3813, 2014.

\bibitem[Border and Sobel(1987)]{border1987samurai}
K.~C. Border and J.~Sobel.
\newblock Samurai accountant: A theory of auditing and plunder.
\newblock \emph{The Review of Economic Studies}, 54\penalty0 (4):\penalty0
  525--540, 1987.

\bibitem[Bouwhuis et~al.(2026)Bouwhuis, Borm, and
  Hendrickx]{bouwhuis2026strategic}
D.~Bouwhuis, P.~Borm, and R.~Hendrickx.
\newblock A strategic approach to bankruptcy problems based on the {TAL} family
  of rules.
\newblock \emph{Annals of Operations Research}, 356:\penalty0 85--104, 2026.

\bibitem[Chun(1989)]{chun1989noncooperative}
Y.~Chun.
\newblock A noncooperative justification for egalitarian surplus sharing.
\newblock \emph{Mathematical Social Sciences}, 17\penalty0 (3):\penalty0
  245--261, 1989.

\bibitem[Dagan et~al.(1997)Dagan, Serrano, and Volij]{dagan1997noncooperative}
N.~Dagan, R.~Serrano, and O.~Volij.
\newblock A noncooperative view of consistent bankruptcy rules.
\newblock \emph{Games and Economic Behavior}, 18\penalty0 (1):\penalty0 55--72,
  1997.

\bibitem[Dagan et~al.(1999)Dagan, Serrano, and Volij]{dagan1999feasible}
N.~Dagan, R.~Serrano, and O.~Volij.
\newblock Feasible implementation of taxation methods.
\newblock \emph{Review of Economic Design}, 4\penalty0 (1):\penalty0 57--72,
  1999.

\bibitem[de~Frutos(1999)]{defrutos1999coalitional}
M.~A. de~Frutos.
\newblock Coalitional manipulations in a bankruptcy problem.
\newblock \emph{Review of Economic Design}, 4\penalty0 (3):\penalty0 255--272,
  1999.

\bibitem[Estornell et~al.(2021)Estornell, Das, and
  Vorobeychik]{estornell2021incentivizing}
A.~Estornell, S.~Das, and Y.~Vorobeychik.
\newblock Incentivizing truthfulness through audits in strategic
  classification.
\newblock \emph{Proceedings of the AAAI Conference on Artificial Intelligence},
  35\penalty0 (6):\penalty0 5347--5354, 2021.

\bibitem[Hagiwara and Hanato(2021)]{hagiwara2021strategic}
M.~Hagiwara and S.~Hanato.
\newblock A strategic justification of the constrained equal awards rule
  through a procedurally fair multilateral bargaining game.
\newblock \emph{Theory and Decision}, 90:\penalty0 233--243, 2021.

\bibitem[Herrero(2003)]{herrero2003equal}
C.~Herrero.
\newblock Equal awards vs. equal losses: Duality in bankruptcy.
\newblock In M.~R. Sertel and S.~Koray, editors, \emph{Advances in Economic
  Design}, pages 413--426. Springer, Berlin, Heidelberg, 2003.

\bibitem[Ju(2003)]{ju2003manipulation}
B.-G. Ju.
\newblock Manipulation via merging and splitting in claims problems.
\newblock \emph{Review of Economic Design}, 8:\penalty0 205--215, 2003.

\bibitem[Ju et~al.(2007)Ju, Miyagawa, and Sakai]{ju2007nonmanipulable}
B.-G. Ju, E.~Miyagawa, and T.~Sakai.
\newblock Non-manipulable division rules in claim problems and generalizations.
\newblock \emph{Journal of Economic Theory}, 132\penalty0 (1):\penalty0 1--26,
  2007.

\bibitem[Kartik(2009)]{kartik2009strategic}
N.~Kartik.
\newblock Strategic communication with lying costs.
\newblock \emph{The Review of Economic Studies}, 76\penalty0 (4):\penalty0
  1359--1395, 2009.

\bibitem[Li and Ju(2023)]{li2023divide}
J.~Li and Y.~Ju.
\newblock Divide and choose: An informationally robust strategic approach to
  bankruptcy problems.
\newblock \emph{Journal of Mathematical Economics}, 107:\penalty0 102862, 2023.

\bibitem[Moreno-Ternero(2007)]{moreno2007manipulation}
J.~D. Moreno-Ternero.
\newblock Bankruptcy rules and coalitional manipulation.
\newblock \emph{International Game Theory Review}, 9\penalty0 (1):\penalty0
  105--118, 2007.
\newblock Erratum: 9(2):411--424.

\bibitem[Moreno-Ternero and Villar(2004)]{moreno2004securement}
J.~D. Moreno-Ternero and A.~Villar.
\newblock The {Talmud} rule and the securement of agents' awards.
\newblock \emph{Mathematical Social Sciences}, 47\penalty0 (2):\penalty0
  245--257, 2004.

\bibitem[Moreno-Ternero and Villar(2006)]{moreno2006tal}
J.~D. Moreno-Ternero and A.~Villar.
\newblock The {TAL}-family of rules for bankruptcy problems.
\newblock \emph{Social Choice and Welfare}, 27\penalty0 (2):\penalty0 231--249,
  2006.

\bibitem[Moreno-Ternero et~al.(2022)Moreno-Ternero, Tsay, and
  Yeh]{moreno2022strategic}
J.~D. Moreno-Ternero, M.-H. Tsay, and C.-H. Yeh.
\newblock Strategic justifications of the {TAL} family of rules for bankruptcy
  problems.
\newblock \emph{International Journal of Economic Theory}, 18\penalty0
  (1):\penalty0 92--102, 2022.

\bibitem[O'Neill(1982)]{oneill1982problem}
B.~O'Neill.
\newblock A problem of rights arbitration from the {Talmud}.
\newblock \emph{Mathematical Social Sciences}, 2\penalty0 (4):\penalty0
  345--371, 1982.

\bibitem[Peters et~al.(2019)Peters, Schr{\"o}der, and
  Vermeulen]{peters2019claim}
H.~Peters, M.~Schr{\"o}der, and D.~Vermeulen.
\newblock Claim games for estate division problems.
\newblock \emph{Games and Economic Behavior}, 116:\penalty0 105--115, 2019.

\bibitem[Serrano(1995)]{serrano1995strategic}
R.~Serrano.
\newblock Strategic bargaining, surplus sharing problems and the nucleolus.
\newblock \emph{Journal of Mathematical Economics}, 24\penalty0 (4):\penalty0
  319--329, 1995.

\bibitem[Sertel(1992)]{sertel1992nash}
M.~R. Sertel.
\newblock The {Nash} bargaining solution manipulated by pre-donations is
  {Talmudic}.
\newblock \emph{Economics Letters}, 40\penalty0 (1):\penalty0 45--55, 1992.

\bibitem[Thomson(2019)]{thomson2019how}
W.~Thomson.
\newblock \emph{How to divide when there isn't enough}.
\newblock Cambridge University Press, 2019.
\newblock ISBN 9781316646441.

\bibitem[Townsend(1979)]{townsend1979costly}
R.~M. Townsend.
\newblock Optimal contracts and competitive markets with costly state
  verification.
\newblock \emph{Journal of Economic Theory}, 21\penalty0 (2):\penalty0
  265--293, 1979.

\bibitem[Tsay and Yeh(2019)]{tsay2019relations}
M.-H. Tsay and C.-H. Yeh.
\newblock Relations among the central rules in bankruptcy problems: A strategic
  perspective.
\newblock \emph{Games and Economic Behavior}, 113:\penalty0 515--532, 2019.

\end{thebibliography}
}

\end{document}